\documentclass[pra,twocolumn]{revtex4-1}
\usepackage{hyperref} 

\usepackage{graphicx} 
\usepackage{subfig}
\usepackage{color}
\usepackage{filecontents}
\usepackage{soul}
\usepackage{bbold}

\newcommand{\ket}[1]{\left| #1 \right\rangle}
\newcommand{\bra}[1]{\left\langle #1 \right|}

\newcommand{\be}{\begin{equation}}
\newcommand{\ee}{\end{equation}}
\newcommand{\bea}{\begin{eqnarray}}
\newcommand{\eea}{\end{eqnarray}}
\newcommand*{\myeqref}[2][Eq.~]{%
  \hyperref[{#2}]{#1(\ref*{#2})}%
}
\def\equationautorefname#1#2\null{%
  Eq.#1(#2\null)%
}

\def\overbigdot#1{\overset{\hbox{\tiny$\bullet$}}{#1}}

\usepackage{dcolumn}
\usepackage{bm}
\usepackage{amssymb,amsmath}
\usepackage{color}
\usepackage{float}
\usepackage{tikz}
\usetikzlibrary{arrows}

\definecolor{DarkGreen}{rgb}{0,0.6,0.2}

\begin{document}
\title{On the dissipative dynamics of entangled states in coupled-cavity quantum electrodynamics arrays}
\author{Imran M. Mirza, and Adriana S. Cruz}
\affiliation{Macklin Quantum Information Sciences, \\
Department of Physics, Miami University, Oxford, Ohio 45056, USA}
\email{mirzaim@miamioh.edu}


\begin{abstract}
We examine the dissipative dynamics of N00N states with an arbitrary photon number $\mathcal{N}$ in two architectures of fiber-coupled optical ring resonators (RRs) interacting with two-level quantum emitters. One architecture consists of a two-way cascaded array of emitter-cavity systems, while in the other architecture we consider two fiber-coupled RRs each coupled to multiple dipole-dipole interacting (DDI) quantum emitters (QEs). Our focus in this paper is to study how am initially prepared multiple excitation atomic N00N states transfers to the RRs and then how rapidly it decays in these open cavity quantum electrodynamics (CQED) setups while varying the emitter-cavity coupling strengths, emitter-cavity detuning, and backscattering from cavity modes. We present a general theoretical formalism valid for any arbitrary numbers of QEs, RRs, and $\mathcal{N}$ number in the N00N state for both schemes. As examples, we discuss the cases of single and two-excitation N00N states and report the comparison of our findings in both schemes. As one of the main results, we conclude that the array scheme tends to store N00N for longer times while the DDI scheme supports higher fidelity values. The results of this study may find applications in designing new multiparty entanglement-based protocols in quantum metrology and quantum lithography.
\end{abstract}


\maketitle
\section{Introduction}
Nonclassical states of light and matter play a decisive role in the processing of information in a quantum-mechanical manner. Despite a broad range of proven and proposed applications in quantum information processing \cite{nielsen2010quantum, bouwmeester2000physics}, quantum computation \cite{bennett2000quantum} and quantum metrology  \cite{giovannetti2006quantum}, these nonclassical states exhibit an extremely fragile behavior when exposed to the surrounding environment. This openness to the outside environment results in a fast decay and (in most of the situations) in an irrecoverable loss of information stored in these states. Among disparate types of nonclassical states, N00N states \cite{boto2000quantum,sanders1989quantum} represents a unique kind of genuine many-body entangled state with two superimposed orthogonal components. The N00N state is generally expressed as
\begin{equation}
\ket{\Psi}=\frac{1}{\sqrt{2}}\left(\ket{\mathcal{N},0}+e^{i\varphi}\ket{0,\mathcal{N}}\right),
\end{equation}
where $0$ and $\mathcal{N}$ represent zero and $N$ number of excitations in the N00N state, and $\varphi$ is a relative phase between two orthogonal state components. 

Remarkable applications of N00N states have already been found in quantum optical lithography \cite{d2001two}, quantum metrology \cite{joo2011quantum}, precision measurements \cite{kok2002creation}, multiparticle entanglement based quantum interferometry \cite{hyllus2012fisher} and quantum-enhanced information processing (in general) \cite{gisin2007quantum}.
Consequently, different experimental demonstrations and theoretical proposals are carried out recently for the successful generation of N00N states in diverse systems. Some captivating examples include: Fast generation of photonic N00N states with two artificial atoms in two microwave cavities \cite{su2014fast, merkel2010generation}, three-photon N00N state generation scheme based on spontaneous parametric down-conversion \cite{kim2009three}, creating high fidelity N00N states through mixing classical and quantum light \cite{afek2010high}, production of atomic N00N states via phase super-resolution measurements \cite{chen2010heralded} and realization of mechanical N00N states in microcavity optomechanics \cite{ren2013single}.

Post generation, a crucial question is how fast a multi-photon N00N state decays given a time requirement set by the open quantum information protocol? In this paper, we address this question by analyzing and comparing two setups to transfer the emitter-generated multi-excitation N00N states to RRs (or optical cavities). One scheme is a two-way cascaded Jaynes-Cummings (JC) array \cite{hartmann2006strongly,greentree2006quantum,mirza2013single,mirza2015nonlinear,mirza2015bi} and the other scheme consists of multiple DDI emitters in two fiber-coupled RRs. To the best of our knowledge, such a study using these particular architectures has not been reported before in the present context.

\begin{figure*}\label{Fig1}
\includegraphics[width=6.8in,height=1.2in]{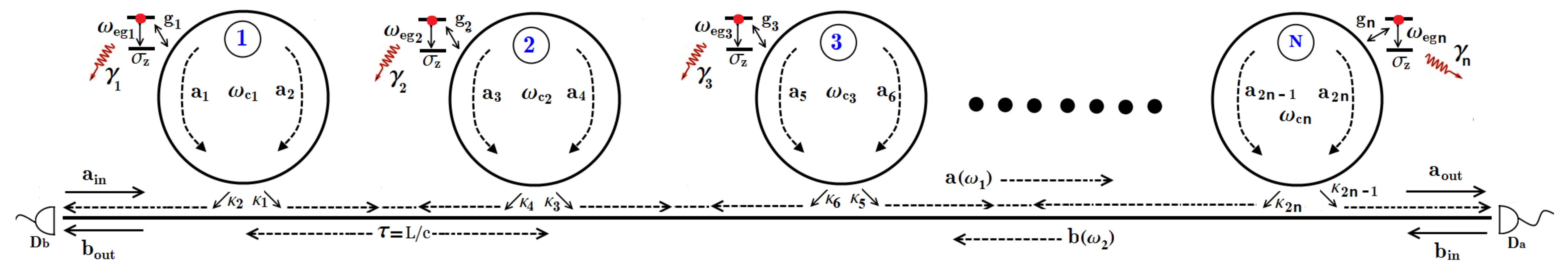}
\captionsetup{
  format=plain,
  margin=1em,
  justification=raggedright,
  singlelinecheck=false
}
\caption{Architecture-I: An array of $N$ number of emitter-cavity systems and two output photodetectors. Due to the two-way cascaded coupling all $\hat{a}_{2n-1}$ (and similarly $\hat{a}_{2n}$) modes can interact with each other in a uni-directional manner while $ n=1,2,...,N$. The fiber time delay between two consecutive emitter-cavity systems is $\tau=L/c$, where L is the distance between two cavities and $c$ is the group velocity of light in the fiber. For simplicity, one can either consider all emitter-cavity systems identical or can assume a mirror-symmetric situation (mirror lying in the middle of the array with an even number of emitter-cavity subsystems). The detectors count photons in the two output modes, described by the continua annihilation operators $\hat{a}_{out}$ and $\hat{b}_{out}$. For further details regarding the system see Sec.~II(A).}\label{Fig1}
\end{figure*}

In our models, no external source of single-photons is included rather excited atoms coupled with their respective cavities serve as the sources of the single photons. The produced photons are then transferred from one emitter-cavity system to the other through the fiber route. For realistic treatment, we have included the effects of different decoherence mechanisms (including the photon leakage from the cavity mirrors and spontaneous emission from QEs) on the time evolution of obtained N00N states. To this end, a quantum jump approach (QJA) combined with the input-output theory \cite{gardiner1985input} for cascaded quantum systems is employed to incorporate uni-directional coupling between the consecutive cavity modes. Furthermore, we utilize the quantum state fidelity \cite{nielsen2010quantum} as a measure to keep track of the decay of the generated N00N state.

We find that in general in the JC array scheme one can sustain the N00N states for longer times, whereas in DDI emitters coupled to the RRs scheme one can obtain the N00N state with higher fidelity. In both schemes, we find that under the strong emitter-cavity coupling fidelity manifests oscillatory behavior which originates from the Rabi oscillations. We also notice with an increase in the excitation number from $\mathcal{N}=1$ to $\mathcal{N}=2$, maximum fidelity achieved decreases markedly in both schemes (range lies between 50\% to almost an order of magnitude decrease).
  
The paper is organized as follows. In Sec.~II, we outline the model of our JC array scheme. We also present the dissipative dynamics in the same section through the quantum Langevin equations \cite{gardiner2004quantum} and the quantum trajectory/jump method \cite{carmichael1993open, daley2014quantum}. In Sec.~III we introduce the setup for scheme-II involving quantum emitters that are directly coupled through the DDI. Next, in Sec.~IV we report the main findings of our study by comparing the time-evolution and maximum fidelity achieved for the uni- and bi-photon N00N states in both schemes. Finally, in Section V, we close with a summary of the main findings of this work.


\section{Architecture-I: JC array}
\subsection{System Hamiltonian and Dissipative Dynamics}
The system under consideration is shown in Fig.~\ref{Fig1}. Under the rotating wave approximation, dipole approximation, and Markov approximation (collectively referred to as the quantum white-noise limit \cite{baragiola2012n}), the Hamiltonian for the global system (system, bath, and system-bath interaction) is expressed as
 \begin{widetext}
\begin{equation}\label{H}
\begin{split}
& \hat{H}=\hbar\sum^{N}_{n=1}\left[-\omega_{egn}\hat{\sigma}_{n}\hat{\sigma}^{\dagger}_{n}+ 
\omega_{cn}\hat{a}_{n}^{\dagger}\hat{a}_{n}\right]+\hbar\sum^{N}_{n=1}\left[g_{n}\hat{a}_{2n-1}^{\dagger}\hat{\sigma}_{n}+g^{\ast}_{n}\hat{a}_{2n-1}\hat{\sigma}^{\dagger}_{n}\right]+\hbar\sum^{N}_{n=1}\left[g^{\ast}_{n}\hat{a}_{2n}^{\dagger}\hat{\sigma}_{n}+g_{n}\hat{a}_{2n}\hat{\sigma}^{\dagger}_{n})\right]\\
&+\hbar\sum^{N}_{n=1}\eta \left(\hat{a}^{\dagger}_{2n-1}\hat{a}_{2n}+\hat{a}^{\dagger}_{2n}\hat{a}_{2n-1}\right)+\hbar\int_{-\infty}^{+\infty}\omega_{1} \hat{a}^\dagger(\omega_{1})\hat{a}(\omega_{1})d\omega_{1}
+\hbar\int_{-\infty}^{+\infty}\omega_{2} \hat{b}^\dagger(\omega_{2})\hat{b}(\omega_{2})d\omega_{2}\\
&+i\hbar\sum^{N}_{n=1}\sqrt{\frac{\kappa_{2n-1}}{2\pi}}\int_{-\infty}^{+\infty}\left(\hat{a}_{2n-1}\hat{a}^{\dagger}(\omega_{1})-\hat{a}_{2n-1}^{\dagger}\hat{a}(\omega_{1})\right)d\omega_{1}+i\hbar\sum^{N}_{n=1}\sqrt{\frac{\kappa_{2n}}{2\pi}}\int_{-\infty}^{+\infty}\left(\hat{a}_{2n}\hat{b}^{\dagger}(\omega_{2})-\hat{a}_{2n}^{\dagger}\hat{b}(\omega_{2})\right)d\omega_{2}.
\end{split}
\end{equation}
\end{widetext}
The setup consists of $N$ number of emitter-cavity subsystems coupled through a dispersion-less optical fiber. QEs have a two-level energy structure with $n$th emitter transition frequency and spontaneous emission rate given by $\omega_{egn}$ and $\gamma_{n}$, respectively. Raising (lowering) of $n$th emitter is described by the atomic raising (lowering) operator $\hat{\sigma}^{\dagger}_{n} (\hat{\sigma}_{n})$. The coupling between the $n$th emitter and the $n$th cavity is quantified through the emitter-cavity coupling rate $g_{n}$ while each $n$th cavity is assumed to support a single isolated resonant mode with frequency $\omega_{cn}$. When the $n$th emitter de-excites and a photon is emitted into the cavity, any one of the two counter-propagating modes (described by annihilation operators $\hat{a}_{2n-1}$ and $\hat{a}_{2n}$) can be excited. Due to cavity modes back reflections, the modes $\hat{a}_{2n-1}$ and $\hat{a}_{2n}$ can directly mix --- a process characterized by the backscattering rate $\eta$. Once a cavity is populated with a photon, it can leak the photon into the tapered fiber, and from the $2n-1$ to the $2n$th cavity modes this leakage is described in terms of rates $\kappa_{2n-1}$ and $\kappa_{2n}$, respectively. Dispersion-less fiber is modeled to have two continua of modes. The annihilation of photons in the left (right) direction continuum is described by the operator $\hat{b}(\omega_{2})$ ($\hat{a}(\omega_{1})$) as shown in Fig.~\ref{Fig1}.

The first and second terms on the right-hand side of Eq.~\eqref{H} represent the free Hamiltonian for the QEs and cavity modes, respectively. Here we have chosen the energy of the emitters' ground states to be negative, such that the initial state has zero energy. We have also neglected the zero-point energy in the cavity mode free Hamiltonian. Next terms with prefactors $g_n$ and $g^\ast_n$ are the emitter-cavity interaction Hamiltonian that is following the standard Jaynes-Cummings model. The first term on the second line of Eq.~\eqref{H} shows the cavity backscattering Hamiltonian. The last two terms on the same line are the free bath Hamiltonians while the terms on the last line present the coupling between cavity modes with their respective baths. The nonvanishing commutation and anticommutation relations among system and bath operators are given by:
\begin{align}
&\lbrace\hat{\sigma}_{n},\hat{\sigma}^{\dagger}_{n^{'}}\rbrace=\delta_{nn^{'}},~~ [\hat {a}(\omega_{i}),\hat {a}^{\dagger}(\omega_{j})]=\delta(\omega_{i}-\omega_{j})\nonumber,\\
& [\hat {b}(\omega_{i}),\hat {b}^{\dagger}(\omega_{j})]=\delta(\omega_{i}-\omega_{j}),~~\text{and}~~ [\hat {a}_{n},\hat {a}_{n^{'}}^{\dagger}]=\delta_{nn^{'}}.
\end{align}
The imposition of the time-reversal symmetry on Hamiltonian requires all cavity decay rates to be identical (i.e. $\kappa_{1}=\kappa_{2}=...=\kappa_{2n}\equiv\kappa$). But here we still use different subscripted cavity decay rates, just to keep track of different coupling terms. Later in plots, one can further simplify the situation by taking all emitter-cavity subsystems to be identical. 


\subsection{Bi-directional quantum Langevin equation and input-output relations}
The interaction of the emitters and intracavity modes with the environment makes our emitter-cavity array essentially an open quantum system. To describe the dynamics of such a quantum system we start off by working in the Heisenberg picture. Therein, by following the standard procedure, \cite{gardiner2004quantum, carmichael1993open} of eliminating continua from the system dynamics, we can identify the two input and output operators \cite{gardiner1985input} corresponding to two fiber continua as
\begin{subequations}
\begin{eqnarray}
\hat{a}_{in}(t):=\frac{1}{\sqrt{2\pi}}\int_{-\infty}^{\infty}\hat{a}_{0}(\omega_{1})e^{-i\omega_{1}(t-t_{0})}d\omega_{1},\\
\hat{b}_{in}(t):=\frac{1}{\sqrt{2\pi}}\int_{-\infty}^{\infty}\hat{b}_{0}(\omega_{2})e^{-i\omega_{2}(t-t_{0})}d\omega_{2}.
\end{eqnarray}
\end{subequations}
\begin{subequations}
\begin{eqnarray}
\hat{a}_{out}(t):=\frac{1}{\sqrt{2\pi}}\int_{-\infty}^{\infty}\hat{a}_{1}(\omega_{1})e^{-i\omega_{1}(t-t_{1})}d\omega_{1},\\
\hat{b}_{out}(t):=\frac{1}{\sqrt{2\pi}}\int_{-\infty}^{\infty}\hat{b}_{1}(\omega_{2})e^{-i\omega_{2}(t-t_{1})}d\omega_{2},
\end{eqnarray}
\end{subequations}
where $t_{0}$ and $t_{1}$ are some arbitrary initial and final times, with  $\hat{a}_{0}(\omega_{1})\equiv\hat{a}(\omega_{1};t_{0}), \hat{b}_{0}(\omega_{1})\equiv\hat{b}(\omega_{1};t_{0}), \hat{a}_{1}(\omega_{1})\equiv\hat{a}(\omega_{1};t_{1}),$ and $ \hat{b}_{1}(\omega_{1})\equiv\hat{b}(\omega_{1};t_{1})$ being the past and future time conditioned continua operators. The temporal evolution of any arbitrary system operator $\hat{X}$ (which may belong to any one of the emitter-cavity systems in the array) can then be expressed through the following quantum Langevin's equation
\begin{widetext}
\begin{equation}\label{Lang}
\begin{split}
& \frac{d\hat{X}(t)}{dt}=-\frac{i}{\hbar}[\hat{X}(t),\hat{H}_{sys}]
-\sum_{n=1}^{N}\Bigg\lbrace\left[\hat{X}(t),\hat{a}_{2n-1}^\dagger\right]\left(\frac{\kappa_{2n-1}}{2}\hat{a}_{2n-1}+\sqrt{\kappa_{2n-1}}\hat{a}_{in}(t-(n-1)\tau)\right)
-\Big(\frac{\kappa_{2n-1}}{2}\hat{a}_{2n-1}^{\dagger}+\sqrt{\kappa_{2n-1}}\\
&\times\hat{a}_{in}^{\dagger}(t-(n-1)\tau)\Big)
\left[\hat{X}^\dagger(t),\hat{a}_{2n-1}\right]\Bigg\rbrace-\sum^{N-1}_{n=1}\sum^{N-1}_{m=1}\Bigg\lbrace\sqrt{\kappa_{2n-1}\kappa_{2m+1}}~\delta_{m>n-1}\left[\hat{X}(t),\hat{a}_{2m+1}^{\dagger}\right]\hat{a}_{2n-1}(t-(n-m+1)\tau)\\
&-\sqrt{\kappa_{2n-1}\kappa_{2m+1}}~\delta_{m>n-1}\hat{a}_{2n-1}^{\dagger}(t-(n-m+1)\tau)\left[\hat{X}(t),\hat{a}_{2m+1}\right]\Bigg\rbrace-\sum_{n=1}^{N}\Bigg\lbrace\left[\hat{X}(t),\hat{a}_{2n}^\dagger\right]\Big(\frac{\kappa_{2n}}{2}\hat{a}_{2n}+\sqrt{\kappa_{2n}}\\
&\times\hat{b}_{in}(t-(N-n)\tau)\Big)-\Big(\frac{\kappa_{2n}}{2}\hat{a}_{2n}^{\dagger}+\sqrt{\kappa_{2n}}\hat{b}_{in}^{\dagger}(t-(N-n)\tau)\Big)\left[\hat{X}^\dagger(t),\hat{a}_{2n}\right]\Bigg\rbrace-\sum^{N-1}_{n=1}\sum^{N-1}_{m=1}\Bigg\lbrace\sqrt{\kappa_{2n}\kappa_{2m+2}}~\delta_{m<n-1}\\
&\times\left[\hat{X}(t),\hat{a}_{2m+2}^{\dagger}\right]\hat{a}_{2n}(t-(n-m+1)\tau)-\sqrt{\kappa_{2n}\kappa_{2m+2}}~\delta_{m<n-1}\hat{a}_{2n}^{\dagger}(t-(n-m+1)\tau)\left[\hat{X}(t),\hat{a}_{2m+2}\right]\Bigg\rbrace.
\end{split}
\end{equation}
\end{widetext}
Here $\hat{H}_{sys}$ is the atom-cavity system Hamiltonian, which consists of the discrete terms in Eq.~[\ref{H}] with $\delta_{m\lessgtr n-1}=1 \forall m\lessgtr n-1$ and otherwise zero. The above Langevin equation is a generalization of the usual cascaded quantum system Langevin equation \cite{gardiner2004quantum} where we have included a bidirectional coupling among atom-cavity systems in the array. The JC array is a cascaded quantum network in which the input of one cavity (in one direction) essentially serves as a time delayed output from the other (nearest neighbor) cavity i.e. for any arbitrary $n$th cavity $\hat{a}^{(n)}_{in}(t)=\hat{a}^{(n-1)}_{out}(t-\tau)$ and similarly for opposite direction: $\hat{b}^{(n-1)}_{in}(t)=\hat{b}^{(n)}_{out}(t-\tau)$.

Corresponding to two input field operators $\hat{a}_{in}(t)$, $\hat{b}_{in}(t)$  appearing in the above Langevin equation there are two output operators $\hat{a}_{out}(t)$, $\hat{b}_{out}(t)$ which are related to the time-delayed input and intra cavity field operators through the following input-output relations \cite{carmichael2009statistical,gardiner1985input,carmichael1993open} as
\begin{align}
\hat{a}_{out}(t)=& ~\hat{a}_{in}(t-N\tau)+\sqrt{\kappa_{2n-1}}\hat{a}_{2n-1}(t)\nonumber\\ 
&+\sqrt{\kappa_{2n-3}}\hat{a}_{2n-3}(t-\tau)+...+\sqrt{\kappa_{1}}\hat{a}_{1}(t-N\tau),\\
\hat{b}_{out}(t)=&~\hat{b}_{in}(t)+\sqrt{\kappa_{2}}\hat{a}_{2}(t)\nonumber\\ 
&+\sqrt{\kappa_{4}}\hat{a}_{4}(t-\tau)+...+\sqrt{\kappa_{2n}}\hat{a}_{2n}(t-N\tau).
\end{align}
From now on, we are going to ignore the time retardation/delays assuming that the time scale of system dynamics is much slower compared to the time required for the photons to propagate from one RR to another RR through the optical fiber. Mathematically, this will mean to work in a regime where $\kappa,g,\eta >>\tau^{-1} (=L/c)$. Non-vanishing commutation relations among input and output operators follow causality conditions i.e.
\begin{align}
[\hat{a}_{{\rm in}}(t),\hat{a}_{{\rm in}}^{\dagger}(t')]=\delta(t-t'), ~[\hat{a}_{{\rm out}}(t),\hat{a}_{{\rm out}}^{\dagger}(t')]=\delta(t-t'),
\end{align}
and similarly for $\hat{b}_{in}(t)$ and $\hat{b}_{out}(t)$ operators.  Moreover, throughout this work, we are going to assume that initially there are no photons in the fiber such that in the expectation values of all relevant normally ordered observables, input operators do not contribute and thus can be neglected altogether.


\subsection{Cascaded master equation and the quantum trajectory analysis}
Transforming now to the Schr\"odinger picture with an arbitrary density operator $\hat{\rho}(t)$ and exploiting the trick: $ {\rm d\langle \hat{X}(t)\rangle/dt=Tr\lbrace\hat{X}(t_{0})d\hat{\rho}(t)/dt\rbrace}$ along with the cyclic property of trace (${\rm Tr\lbrace...\rbrace}$) operation, one can drive the master equation in the Lindblad form using the Langevin equation mentioned in Eq.~(\ref{Lang}). The resultant master equation describing the time evolution of system density operator $\hat{\rho}_{s}(t)$ takes the following form
\begin{widetext}
\begin{equation}\label{mas-(1)}
\begin{split}
&\frac{d\hat{\rho}_s(t)}{dt}=\frac{-i}{\hbar}\left[\hat{H}_{sys},\hat{\rho}_s(t)\right]+\sum_{n=1}^{N}\Bigg\lbrace\kappa_{2n-1}\left(\hat{a}_{2n-1}\hat{\rho}_{s}(t)\hat{a}^{\dagger}_{2n-1}-\frac{1}{2}\hat{\rho}_{s}(t)\hat{a}^{\dagger}_{2n-1}\hat{a}_{2n-1}-\frac{1}{2}\hat{a}^{\dagger}_{2n-1}\hat{a}_{2n-1}\hat{\rho}_{s}(t)\right)\Bigg\rbrace \\
&+\sum^{N-1}_{n=1}\sum^{N-1}_{m}\sqrt{\kappa_{2n-1}\kappa_{2m+1}}~\delta_{m>n-1}\left(\left[\hat{a}_{2n-1}\hat{\rho}_{s}(t),\hat{a}^{\dagger}_{2m+1}\right]-\left[\hat{a}_{2m+1},\hat{\rho}_{s}(t)\hat{a}^{\dagger}_{2n-1}\right]\right)+\sum_{n=1}^{N}\Bigg\lbrace\kappa_{2n}\Big(\hat{a}_{2n}\hat{\rho}_{s}(t)\hat{a}^{\dagger}_{2n}\\
&-\frac{1}{2}\hat{\rho}_{s}(t)\hat{a}^{\dagger}_{2n}\hat{a}_{2n}-\frac{1}{2}\hat{a}^{\dagger}_{2n}\hat{a}_{2n}\hat{\rho}_{s}(t)\Big)\Bigg\rbrace+\sum^{N-1}_{n=1}\sum^{N-1}_{m}\sqrt{\kappa_{2n}\kappa_{2m+2}}~\delta_{m<n-1}\left(\left[\hat{a}_{2m+2}\hat{\rho}_{s}(t),\hat{a}^{\dagger}_{2n}\right]-\left[\hat{a}_{2n},\hat{\rho}_{s}(t)\hat{a}^{\dagger}_{2m+2}\right]\right).
\end{split}
\end{equation}
\end{widetext} 
Now to apply the quantum trajectory method (or quantum jump approach (QJA)) \cite{carmichael1993open, molmer1993monte,van2000quantum} which is an appropriate formalism for the description of cascaded open quantum systems, we re-write the above Master equation in a more suggestive form as
\begin{align}
\frac{d\hat{\rho}_s(t)}{dt}&=\frac{-i}{\hbar}\left[\hat{H}^{'},\hat{\rho}_s(t)\right] \nonumber\\
&+\sum_{i=o,e}\left(\hat{J}_{i}\hat{\rho}_s(t)\hat{J}^{\dagger}_{i}-\frac{1}{2}\hat{J}^{\dagger}_{i}\hat{J}_{i}\hat{\rho}_s(t)-\frac{1}{2}\hat{\rho}_s(t)\hat{J}^{\dagger}_{i}\hat{J}_{i}\right),
\end{align}
while $\hat{J}_{e}=\sum^{N}_{n=1}\sqrt{\kappa_{2n}}\hat{a}_{2n}$ and $\hat{J}_{o}=\sum^{N}_{n=1}\sqrt{\kappa_{2n-1}}\hat{a}_{2n-1}$ are the so-called ``jump operators" in the terminology of QJA describing the cascaded coupling among right and left cavity modes, respectively and the Hamiltonian $\hat{H}^{'}$ is given by
\begin{align}
\hat{H}^{'}=&~\hat{H}_{sys}-\frac{i\hbar}{2}\sum^{N-1}_{n=1}\sum^{N-1}_{m=1}\sqrt{\kappa_{2n-1}\kappa_{2m+1}}~\delta_{n<m+1}\Big(\hat{a}_{2n-1}\nonumber\\
&\times\hat{a}^{\dagger}_{2m+1}-\hat{a}^{\dagger}_{2n-1}\hat{a}_{2m+1}\Big).
\end{align}
Combining terms we can rewrite the above master equation in a more compact form as
\begin{equation}\label{mas-(1U)}
\frac{d\hat{\rho}_s(t)}{dt}=\frac{-i}{\hbar}\left[\hat{H}_{NH},\hat{\rho}_s(t)\right]+\sum_{i=o,e}\hat{J}_{i}\hat{\rho}_s(t)\hat{J}^{\dagger}_{i}.
\end{equation}
We notice that if we just consider the first term on the right side of the above equation, then it represents a Liouville-von Neumann type equation with the difference that instead of the system Hamiltonian we now have an effective non-Hermitian Hamiltonian ($\hat{H}_{NH}$). In the QJA this non-Hermitian Hamiltonian consists of a Hermitian (system Hamiltonian) part and an anti-Hermitian part constructed from the jump operators. The explicit form of this non-Hermitian Hamiltonian for the system under consideration is given as follows
\begin{align}
\hat{H}_{NH}&=\hat{H}_{sys}-\frac{i\hbar}{2}\sum^{N}_{n=1}\left(\kappa_{2n-1}\hat{a}^{\dagger}_{2n-1}\hat{a}_{2n-1}+\kappa_{2n}\hat{a}^{\dagger}_{2n}\hat{a}_{2n}\right)\nonumber\\
&-i\hbar\sum^{N-1}_{n=1}\sum^{N-1}_{m=1}\sqrt{\kappa_{2n-1}\kappa_{2m+1}}~\delta_{n<m+1}\hat{a}^{\dagger}_{2m+1}\hat{a}_{2n-1}\nonumber\\
&-i\hbar\sum^{N-1}_{n=1}\sum^{N-1}_{m=1}\sqrt{\kappa_{2n}\kappa_{2m+2}}~\delta_{n>m+1}\hat{a}^{\dagger}_{2m}\hat{a}_{2n+2}.
\end{align}
Note that in the above Hamiltonian we have the three types of terms on the right-hand side: the first term is the usual system Hamiltonian, the next two terms describing the decay of energy from the cavity modes and the last two terms are separately expressing a uni-directional coupling among the even and odd cavity modes through the fiber.
\begin{figure*}
\includegraphics[width=6.5in,height=2.65in]{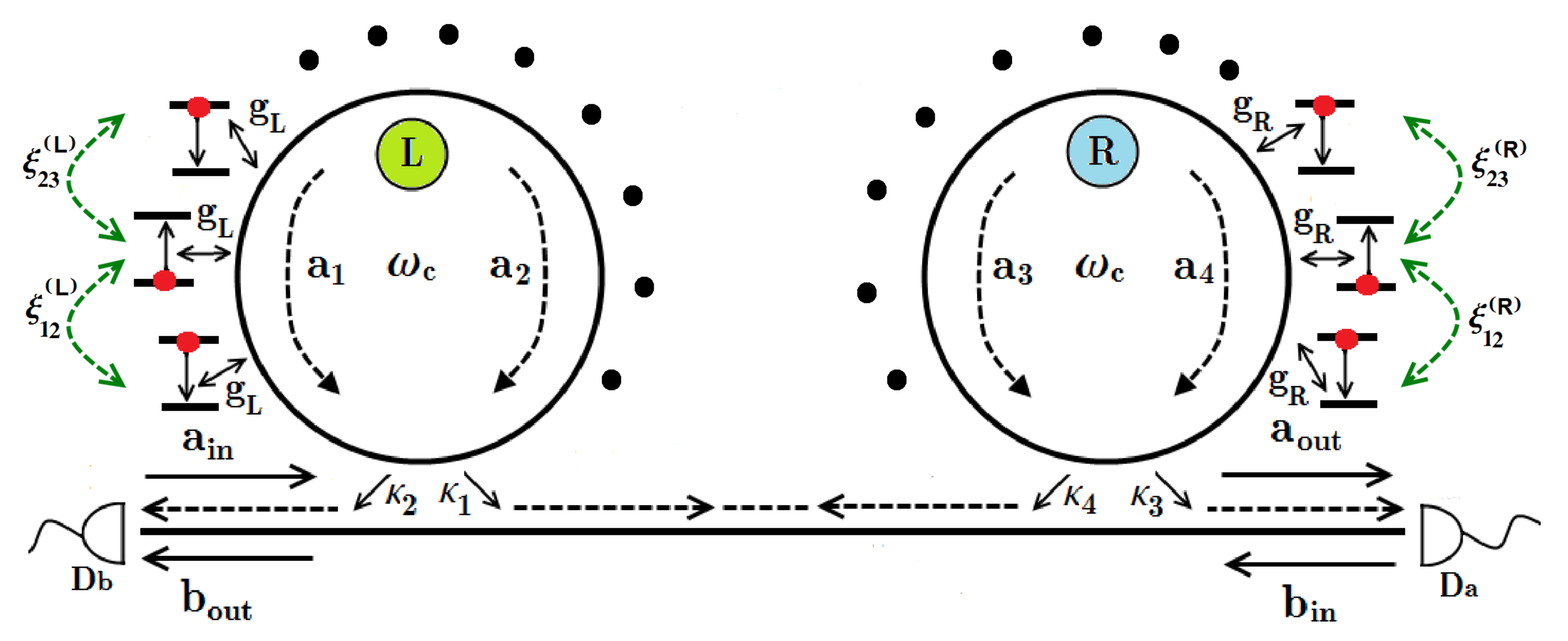}
\captionsetup{
format=plain,
margin=1em,
justification=raggedright,
singlelinecheck=false
}
\caption{Architecture-II: Two fiber-coupled RRs (left or L and right or R), each coupled with $N$ number of QEs positioned on the circumference of the RR and capable of direct coupling through the DDI. Just like scheme-I, the fiber delay has been ignored in the development of mode couplings between the two cavities. Again photo-detectors $D_{a}$ and $D_{b}$ are placed at the ends of the fiber where they can register any number of emitted photons involved in the problem. Black dots in the figure are representing the repetition of the emitters on the RR circumference.}\label{Fig2}
\end{figure*} 
One can then interpret the commutator term appearing in the master equation (Eq.~(\ref{mas-(1U)})) as the system evolution in a non-unitary manner under the action of non-Hermitian Hamiltonian following the so-called non-unitary Schr\"odinger equation in the terminology of the QJA
\begin{equation}\label{NUSE}
i\hbar~\frac{d\ket{\tilde{\psi}(t)}}{dt}=\hat{H}_{NH}{\ket{\tilde{\psi}(t)}}.
\end{equation}
The last term in Eq.~\eqref{mas-(1U)} are the jump terms that describe the decay of energy from the system in a stochastic manner, such that the probability of occurrence of a jump during an infinitesimal time interval $[t,t+dt]$ is given by
\begin{equation}\label{pi}
P_{j}(t)={\bra{\tilde{\psi}}}\hat{J_{j}}^{\dagger}
\hat{J_{j}}{\ket{\tilde{\psi}}}dt=:\Pi_j dt,
\end{equation}
for $j=o,e$ and $\hat{J}_{o}\equiv\hat{a}_{out}(t)$ and $\hat{J}_{e}\equiv\hat{b}_{out}(t)$. After one jump is recorded state of the system is re-normalized (reset) according to the transformation: $\ket{\psi}\mapsto
\frac{\hat{J}_j\ket{\tilde{\psi}}}{\sqrt{\Pi_j}}.$


\section{Architecture-II: DDI emitters in two coupled cavities}
In scheme-II (which is shown in Fig.~\ref{Fig2}), we consider just two fiber-coupled RRs, but each coupled to $N$ number of emitters close enough to be directly coupled to each other through the DDI. The DDI between $n$th and $(n+1)$ QE belonging to $j$th RR is characterized by the parameter $\xi^{(j)}_{n,n+1}$. The absolute position of each atom on the circumference of the RR is not relevant to the present problem, but can easily be included in the model by introducing an appropriate phase in the emitter-cavity coupling rate. Under the rotating wave approximation one can express the Hamiltonian for this scheme as
\begin{widetext}
\begin{equation}
\begin{split}
&\hat{H}_{sys}=-\sum_{j=L,R}\sum^{N}_{n=1}\frac{\hbar\omega_{eg}}{2}\hat{\sigma}^{(j)}_{z,n}+\hbar\omega_{c1}\left(\hat{a}^{\dagger}_{1}\hat{a}_{1}+\hat{a}^{\dagger}_{2}\hat{a}_{2} \right)+\hbar\omega_{c2}\left(\hat{a}^{\dagger}_{3}\hat{a}_{3}+\hat{a}^{\dagger}_{4}\hat{a}_{4} \right)+\sum_{j=L,R}\sum^{N-1}_{n=1}\hbar\xi^{(j)}_{n,n+1}\Big(\hat{\sigma}^{\dagger (j)}_{n}\hat{\sigma}^{(j)}_{n+1} +\\
&+ \hat{\sigma}^{\dagger (j)}_{n+1}\hat{\sigma}^{(j)}_{n}\Big)+\hbar\eta^{(L)}\left(\hat{a}^{\dagger}_{1}\hat{a}_{2}+\hat{a}^{\dagger}_{2}\hat{a}_{1}\right)+\hbar\eta^{(R)}\left(\hat{a}^{\dagger}_{3}\hat{a}_{4}+\hat{a}^{\dagger}_{4}\hat{a}_{3}\right)+\sum^{N}_{n=1}\hbar\left(g_{L}\hat{a}_{1}\hat{\sigma}^{\dagger(L)}_{n} + g^{\ast}_{L}\hat{a}^{\dagger}_{1}\hat{\sigma}^{(L)}_{n}+g^{\ast}_{L}\hat{a}_{2}\hat{\sigma}^{\dagger(L)}_{n} + g_{L}\hat{a}^{\dagger}_{2}\hat{\sigma}^{(L)}_{n} \right)\\
&+\sum^{N}_{n=1}\hbar\left(g_{R}\hat{a}_{3}\hat{\sigma}^{\dagger(R)}_{n} + g^{\ast}_{L}\hat{a}^{\dagger}_{3}\hat{\sigma}^{(R)}_{n}+g^{\ast}_{R}\hat{a}_{4}\hat{\sigma}^{\dagger(R)}_{n} + g_{R}\hat{a}^{\dagger}_{4}\hat{\sigma}^{(R)}_{n} \right).
\end{split}
\end{equation}
\end{widetext}
The dissipative dynamics of the system in this scheme also follow the formalism of the cascaded master equation discussed in detail earlier in the last section. The main difference (from the perspective of the QJA) is now we have just two cavity modes that are cascaded coupled (i.e. $\hat{a}_{1}\leftrightarrow\hat{a}_{3}$ and $\hat{a}_{4}\leftrightarrow\hat{a}_{2}$). Note that in the QJA one can easily incorporate the spontaneous emission rate from the QEs (not drawn in the figure). 

Equipped with this theoretical formalism, In the next section, we'll discuss how one can transfer the atomic excitation-based N00N state to the multiphoton RRs N00N state in both architectures. And we'll also present a comparison (in terms of the cavity N00N state fidelity) between how the generated N00N state evolves in both schemes.


\section{Results and Discussion}
To quantify and examine the time evolution of the obtained state in comparison to the desired multiphoton N00N state, we make use of Uhlamann fidelity \cite{nielsen2010quantum} defined as
\begin{equation}
\mathcal{F}=\Bigg[{\rm Tr}\Bigg
\lbrace\sqrt{\sqrt{\hat{\rho}_{g}}\hat{\rho}_{id}\sqrt{\hat{\rho}_{g}}}\Bigg\rbrace\Bigg]^{2}=\Bigg[\sqrt{\bra{\Psi}\hat{\rho}_{g}\ket{\Psi}}\Bigg]^{2},
\end{equation}
while $\hat{\rho}_{g}$ ($\hat{\rho}_{id}$) is the generated (required/ideal) state. Since our required state is always a N00N state which is a pure state $\ket{\Psi}$ hence above definition simplifies to the last term in the above equation.

Note that the aforementioned open quantum system framework is valid for any number of QEs and RRs in both schemes. We'll focus on applying this framework to two relatively simple examples of single and two-photon N00N states in the following section, where we'll utilize the above-defined fidelity as a quantitative measure to describe N00N state dissipative dynamics.


\subsection{Uni-photon NOON state}
First of all, we'll consider a uni-photon N00N state, which (neglecting the local phase between orthogonal components) takes the form:
\begin{align}
\ket{\Psi}=\frac{1}{\sqrt{2}}(\ket{1,0}+\ket{0,1}).    
\end{align}
This state is one of the four maximally entangled Bell states, which have their own significance due to a wide range of applications in quantum information processing \cite{nielsen2010quantum}, linear optics quantum computation \cite{knill2001scheme}, quantum cryptography \cite{bennett1993teleporting}, quantum dense coding \cite{mattle1996dense} and quantum teleportation \cite{furusawa1998unconditional} (to name a few).

For starting with such a state in emitters, we'll consider just two emitter-cavity subsystems (subsystem (L) and (R)) with one of the emitters being excited initially, but which one we don't know. And then as time evolves such a state can be transferred from emitters to any two of the cavity modes of different emitter-cavity subsystems or even between the hybrid emitter-cavity systems. Note that this setup is valid for both scheme-I and -II. In the next section when we'll consider two excitation cases, then the setups for scheme-I and -II will become different and will be compared. The purpose of this single excitation case is to introduce the notations as well as to study how a single excitation atomic N00N state can be relocated from emitters to cavities and hybrid emitter-cavity subsystems and then can be stored.

Employing the machinery of QJA, for the pre-photodetection times the system under consideration evolves according to the following no-jump state
\begin{align}
\ket{\tilde{\psi}}=&\Bigg(c_{1}(t)\hat{\sigma}^\dagger_1 +c_2(t)\hat{a}^\dagger_1+c_3(t)\hat{a}^\dagger_2+ c_{4}(t)\hat{\sigma}^\dagger_2\nonumber\\
&+c_5(t)\hat{a}^\dagger_3+c_6(t)\hat{a}^\dagger_4\Bigg)\ket{\varnothing},  
\end{align}
where $\ket{\varnothing}$ represents the ground state with both QEs unexcited and no photons in the cavity modes. Inserting the no-jump state along with the non-Hermitian Hamiltonian $\hat{H}_{NH}$ into non-unitary Schr\"odinger equation yields the following set of coupled differential equations for no-jump probability amplitudes
\begin{subequations}
\begin{eqnarray}
\overbigdot{c}_1(t) = -i g^* c_{2}(t)-i g c_{3}(t),~\\
\overbigdot{c}_{2}(t) = -i( \Delta_{ac}-\frac{i}{2} \kappa) c_{2}(t)-i g c_{1}(t),~\\
\overbigdot{c}_{3}(t) = -i(\Delta_{ac}-\frac{i}{2} \kappa) c_{3}(t)- \kappa  c_{6}(t)-i g^* c_{1}(t),~\\
\overbigdot{c}_{4}(t) = -i g^* c_{5}(t)-ig c_{6}(t),~\\
\overbigdot{c}_{5}(t) =  -i(\Delta_{ac}-\frac{i}{2} \kappa) c_{5}(t)-i g c_{4}(t)-i \kappa  c_{2}(t),~\\
\overbigdot{c}_{6}(t) =  -i(\Delta_{ac}-\frac{i}{2} \kappa) c_{6}(t)-i g^* c_{4}(t).~
\end{eqnarray}
\end{subequations}
After finding the solution of the above equation set, one can follow the standard procedure of QJA \cite{di2008photon} to obtain the density operator by performing an ensemble average over the possible realizations of the conditioned output photo-detection. For the present case the density operator can then be  divided into a jump part (with subscript $J$) and a no-jump part (subscript $NJ$) as
\begin{equation}
\hat{\rho}(t)=\mathcal{P}_{NJ}(t)\hat{\rho}_{NJ}(t)+\mathcal{P}_{J}(t)\hat{\rho}_{J}(t).
\end{equation}
For the uni-photon case, only one jump is possible to be recorded hence we can write the density operator more explicitly as
\begin{equation}
\hat{\rho}_s(t)=\ket{\tilde{\psi}}\bra{\tilde{\psi}}+\left|c_{g}(t)\right|^{2}\ket{\varnothing}\bra{\varnothing},
\end{equation}
while we have used: $\hat{\rho}_{NJ}(t)=\frac{\ket{\tilde{\psi}}\bra{\tilde{\psi}}}{||\tilde{\psi}||^{2}}$ and $\mathcal{P}_{NJ}(t)=||\tilde{\psi}||^{2}$ and defined $\mathcal{P}_{J}(t)=|c_{g}(t)|^{2}=1-\sum^{6}_{i=1}|c_{i}(t)|^{2}$. The physical interpretation of $|c_{g}(t)|^{2}$ is the probability of finding the system in the ground state when the single photon is being detected. Note that the definition of $|c_{g}(t)|^{2}$ ensures correct normalization of the obtained density operator.

\begin{figure*}
\begin{center}
\begin{tabular}{cccc}
\subfloat{\includegraphics[width=5.5cm,height=6cm]{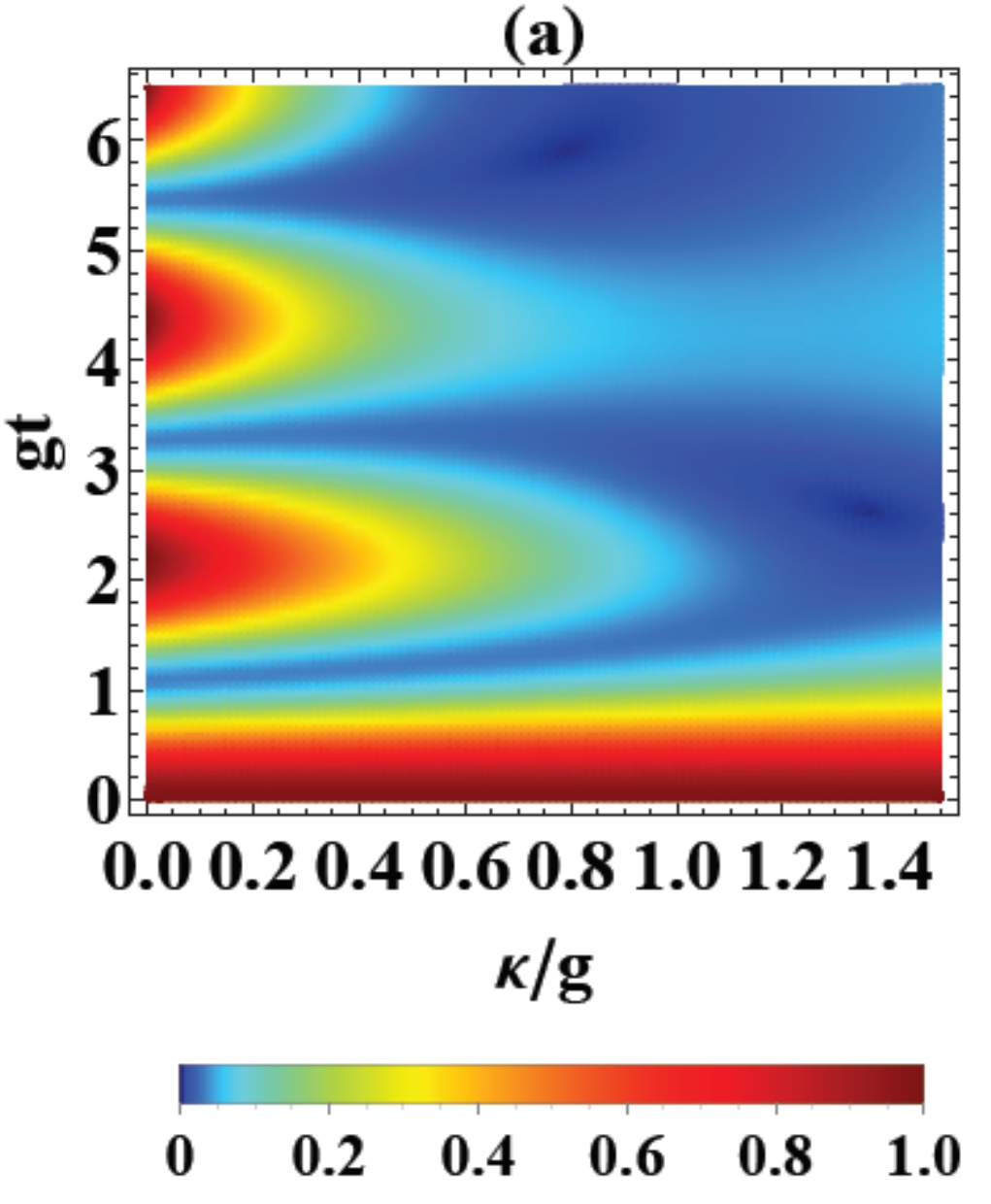}} & 
\subfloat{\includegraphics[width=5.5cm,height=6cm]{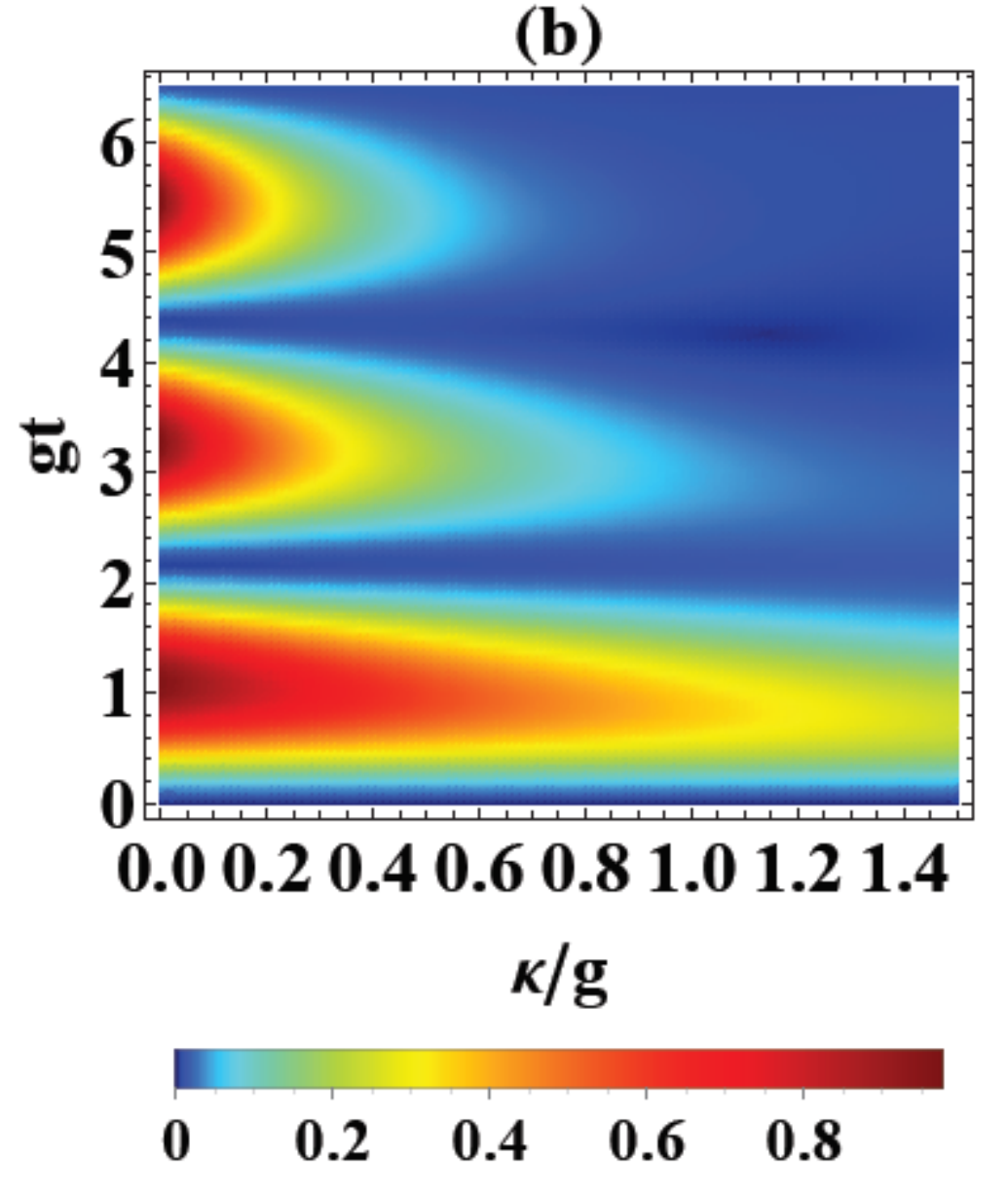}}&
\subfloat{\includegraphics[width=5.5cm,height=6cm]{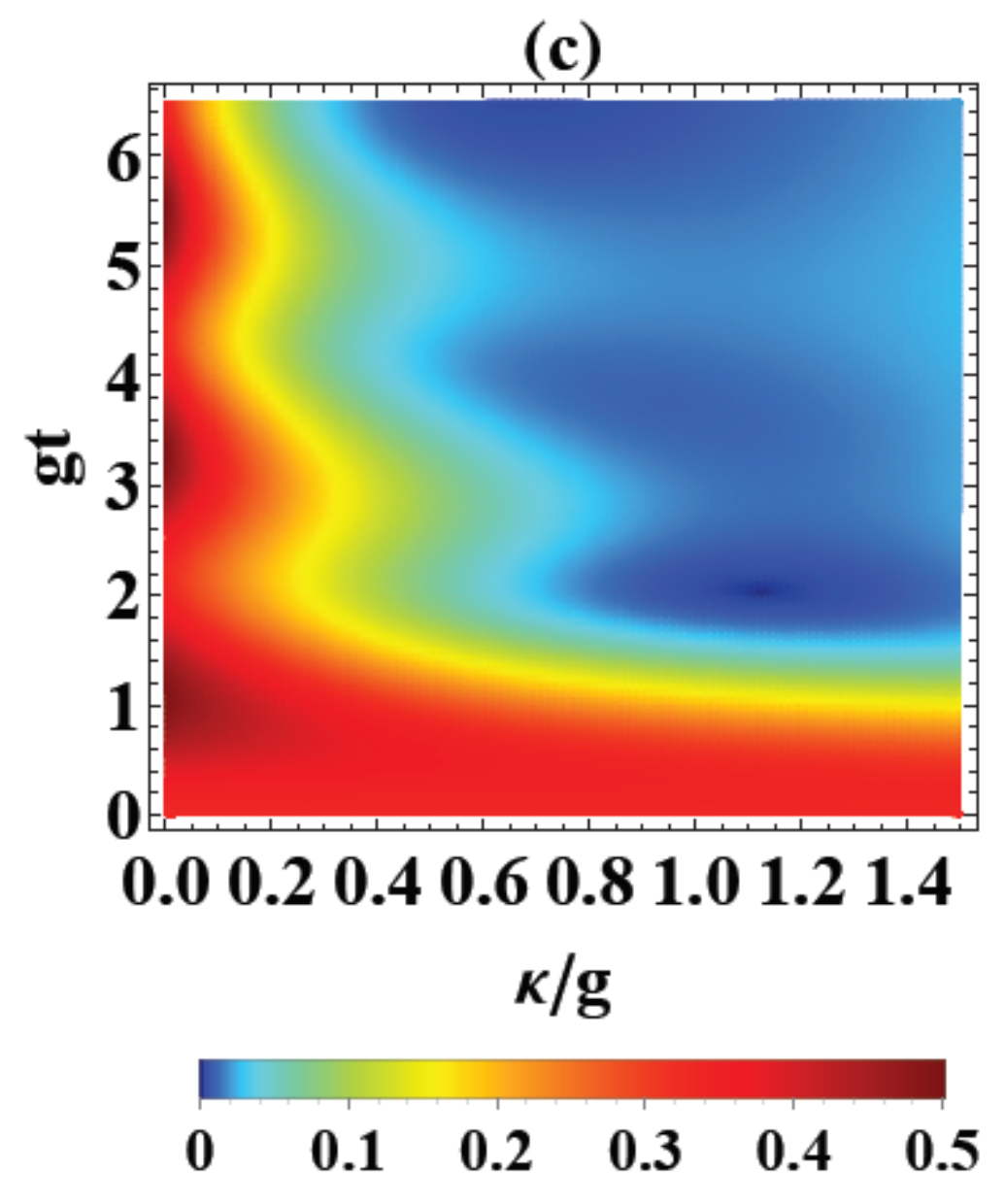}}\\
\end{tabular}
\captionsetup{
  format=plain,
  margin=1em,
  justification=raggedright,
  singlelinecheck=false
}
\caption{Single-photon N00N state dissipative dynamics, when N00N state is prepared between (a) QEs (b) cavity modes (belonging to different cavities) (c) hybrid atom-cavity systems. Parameters used are: $\Delta_{ac}\equiv(\omega_{eg}-\omega_{c})=0.5g$ and the value of cavity decay rate $\kappa$ is varied between 0 and 1.5 in units of emitter-cavity coupling rate $g$ to examine the strong and weak coupling regimes in detail. Note that all decay rates are assumed to be identical here i.e. $\kappa_{1}=\kappa_{2}=...=\kappa_{2n}=\kappa_{2n-1}\equiv\kappa$ with neglecting cavity backscattering ($\eta=0$). Spontaneous emission rates from all QEs have been completely ignored in this and all other plots.}\label{Fig3}
\end{center}
\end{figure*}

Once the full density operator of the system $\hat{\rho}_s(t)$ is attained, we can calculate the time evolution of fidelity of the N00N states. To study how the initial emitter N00N state evolves as a function of time we trace out all four cavity modes from the system density operator $\hat{\rho}_s(t)$ and as a result, obtain the following emitter density operator $\hat{\rho}_E(t)$
\begin{equation}
\begin{split}
&\hat{\rho}_E(t)=Tr_c\lbrace\hat{\rho}_s(t)\rbrace\\
&=|c_{1}(t)|^{2}\ket{e_{1},g_{2}}\bra{e_{1},g_{2}}+|c_{4}(t)|^{2}\ket{g_{1},e_{2}}\bra{g_{1},e_{2}}\\
&+c_{1}(t)c^{\ast}_{4}(t)\ket{e_{1},g_{2}}\bra{g_{1},e_{2}}+c^{\ast}_{1}(t)c_{4}(t)\ket{g_{1},e_{2}}\bra{e_{1},g_{2}}\\
&+\left(1-|c_{1}(t)|^{2}-|c_{4}(t)|^{2}\right)\ket{g_{1},g_{2}}\bra{g_{1},g_{2}}.
\end{split}
\end{equation}
The fidelity ($\mathcal{F}_E(t)$) of such an emitter N00N state (Bell state: $\ket{\Psi}=\frac{1}{\sqrt{2}}(\ket{e_{1},g_{2}}+\ket{g_{1},e_{2}})$) then turns out to be
\begin{equation}
\mathcal{F}_E(t)=\frac{1}{2}\left|c_{1}(t)+c_{4}(t)\right|^{2}.
\end{equation} 
Similarly the fidelity ($\mathcal{F}_c(t)$) of generating single-photon N00N state between cavity modes of left (L) and right (R) cavities ( $\ket{\Psi}=\frac{1}{\sqrt{2}}(\ket{1_{L},0_{R}}+\ket{0_{L},1_{R}})$) takes the form
\begin{equation}
\mathcal{F}_c(t)=\frac{1}{4}|c_{2}(t)+c_{3}(t)+c_{5}(t)+c_{6}(t)|^{2}.
\end{equation}
And finally, the single excitation (emitter excitation or single-photon in any cavity mode) N00N state between hybrid left and right atom-cavity systems follow the fidelity
\begin{equation}
\mathcal{F}_{Ec}(t)=\frac{1}{6}|c_{1}(t)+c_{2}(t)+c_{3}(t)+c_{4}(t)+c_{5}(t)+c_{6}(t)|^{2}.
\end{equation}
Starting with single excitation N00N state in Fig.~3, we plot all three fidelities $\mathcal{F}_{E}(t)$, $\mathcal{F}_{c}(t)$, and $\mathcal{F}_{Ec}(t)$ as a function of time. We notice that in all figures in the strong coupling regime $\kappa<g$ the obtained fidelities oscillate. These are the well-known Rabi oscillations which are describing an almost reversible exchange of photons between the emitter and cavity. Whereas, in the weak coupling regime $\kappa>g$ fidelities exhibit a purely decaying trend. Another noticeable point is that in the initially prepared emitter N00N state while transferring excitations to the cavities the resultant maximum fidelity decreases slightly to 96\% due to the small photon leakage from the imperfect cavity walls. In the cavities, single-photon N00N state tends to stay inside cavity modes for at least $6g^{-1}$ time in the strong coupling regime ($g=10\kappa$) with a 75\% fidelity. Finally, we observe that one can store this single excitation N00N state in the hybrid atom-cavity systems, achieving fidelity of 50\% (as expected) for almost $6g^{-1}$ time.

\begin{figure*}
\begin{center}
\begin{tabular}{cccc}
\subfloat{\includegraphics[width=5.5cm,height=6cm]{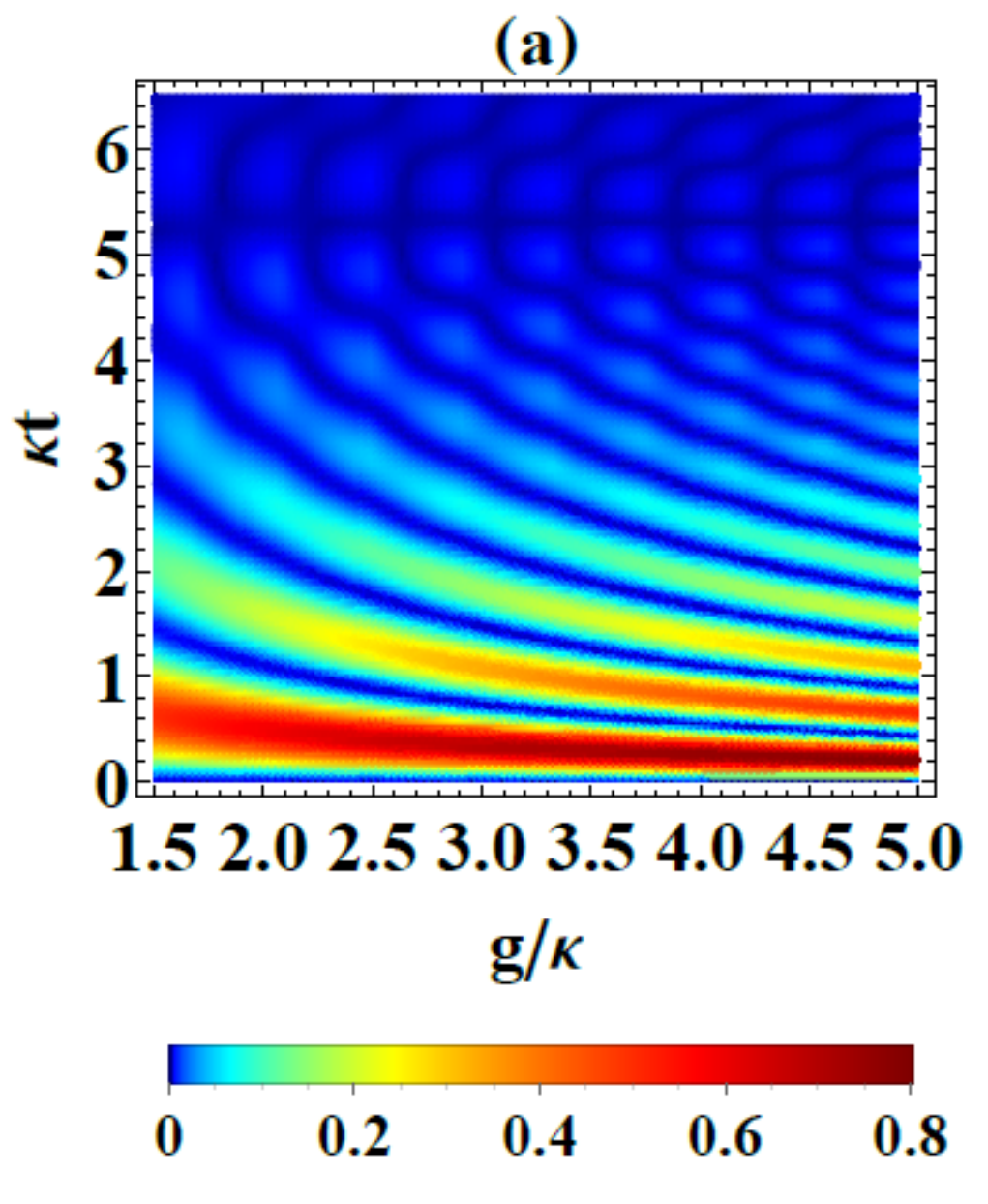}} & 
\subfloat{\includegraphics[width=5.5cm,height=6cm]{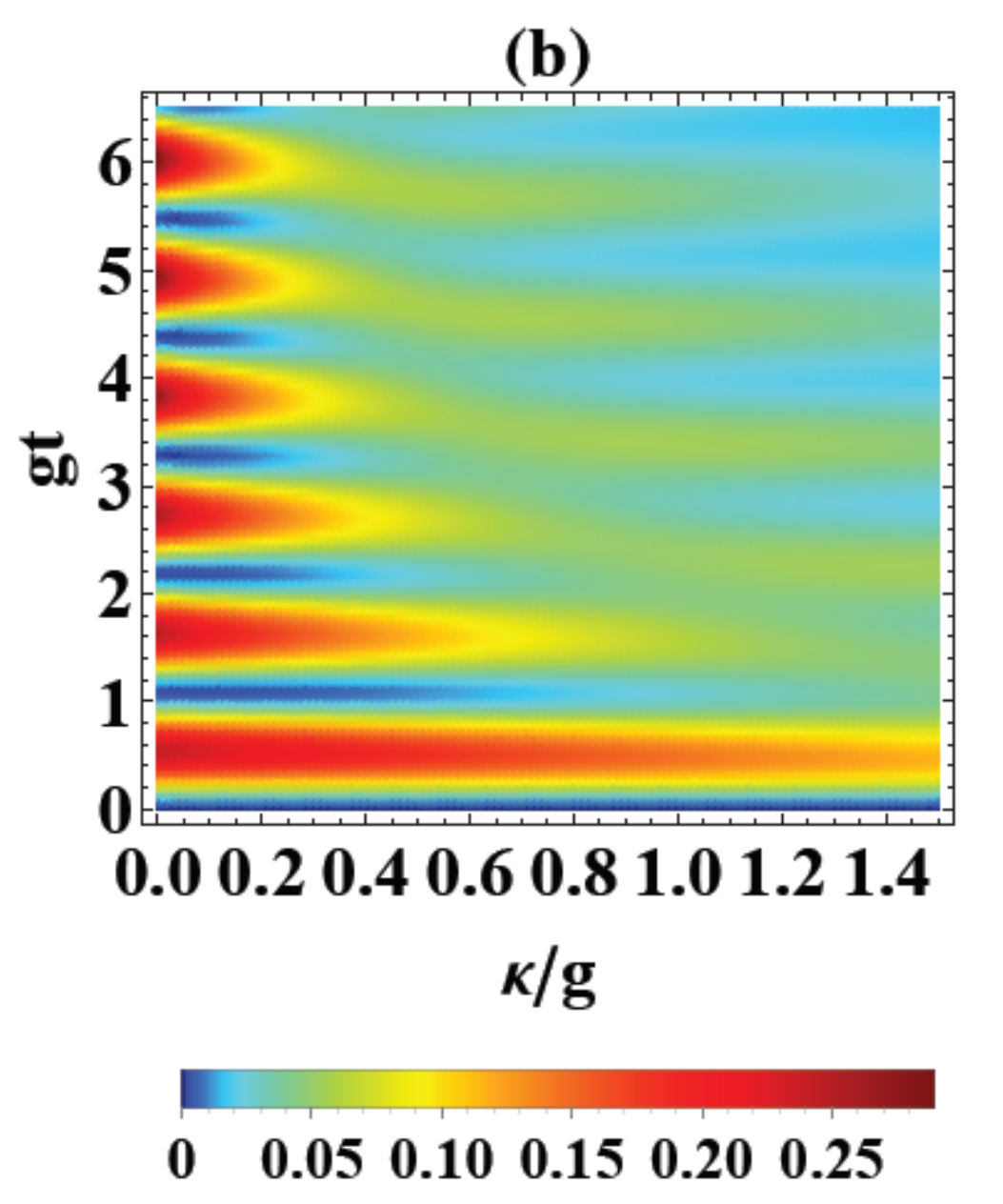}}&
\subfloat{\includegraphics[width=5.5cm,height=6cm]{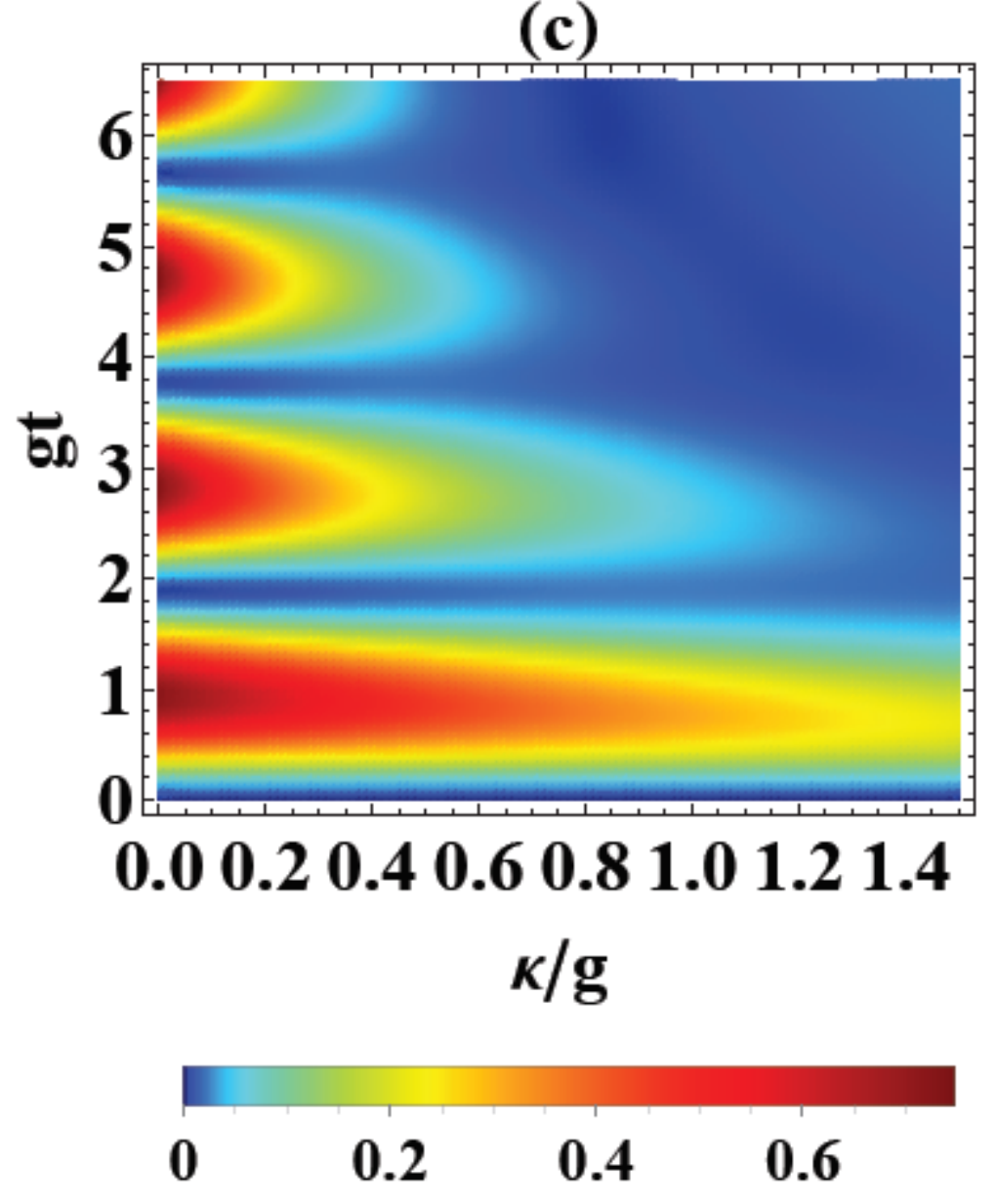}}\\
\end{tabular}
\captionsetup{
  format=plain,
  margin=1em,
  justification=raggedright,
  singlelinecheck=false
}
\caption{Fidelity of the single-photon cavity N00N state under the variation of (a) coupling of emitters with the cavity modes (b) emitter-cavity detuning (far detuned case $\Delta_{ac}=5g$) and (c) strong cavity mode backscattering with $\eta=1.5g$. Similar to Fig.~\ref{Fig3} the value of cavity decay rate $\kappa$ is varied between 0 and 1.5 in units of $g$ in part (b) and (c) of the above figure (to examine the strong and weak coupling regime of CQED in the presence of large detunings and cavity mode backscattering).}\label{Fig4}
\end{center}
\end{figure*}

From now on we concentrate on the cavity N00N state only, i.e. in each case we'll start in an emitter N00N state and then we'll be interested in studying how cavities can store photonic N00N state for protracted times. To this end, in Fig.~4 we have varied the parameter $\kappa/g$, atom-cavity detuning, and cavity backscattering to further examine the dependence of cavity N00N state fidelity on the strong coupling regime of CQED, frequency mismatch, and mixing of modes in each cavity, respectively.

In Fig.~\ref{Fig4}(a) we notice that as we enhance the strong coupling regime, although we obtain a high fidelity  ($\sim$ 80\%) but $\mathcal{F}^{(c)}$ starts to oscillate more rapidly as the period of oscillations depend on the parameter $g$. Due to the strong coupling between emitter and cavity modes and not among cavity modes, the single-photon N00N state vanishes quickly ($\sim t=2\kappa^{-1}$). In all of the previous plots, we were considering a small detuning between the atom and cavity ($\omega_{c}-\omega_{eg}\equiv\Delta_{ac}=0.5g$). In Fig.~ \ref{Fig4}(b) we introduce a large detuning ($\Delta_{ac}=5g$). As a result, we notice that the fidelity remains trapped for longer times (more than $6g^{-1}$) but this happens at the cost of losing fidelity to 25\%. In Fig.~\ref{Fig}(c) we have introduced a strong backscattering ($\eta=1.5 g$) between the cavity modes and as a result, we notice that the highest fidelity achieved reaches $75\%$. Compared to no backscattering situation (Fig.~\ref{Fig3}(b)) we notice the appearance of an additional half blob of fidelity appearing at $\kappa=0$ point. This manifests a control over the collapse and revival pattern of photonic N00N state fidelity due to the possibility of populating both cavity modes by altering the backscattering parameter.


\begin{figure*}
\begin{center}  
\begin{tabular}{cccc}
\subfloat{\includegraphics[width=5.5cm,height=6cm]{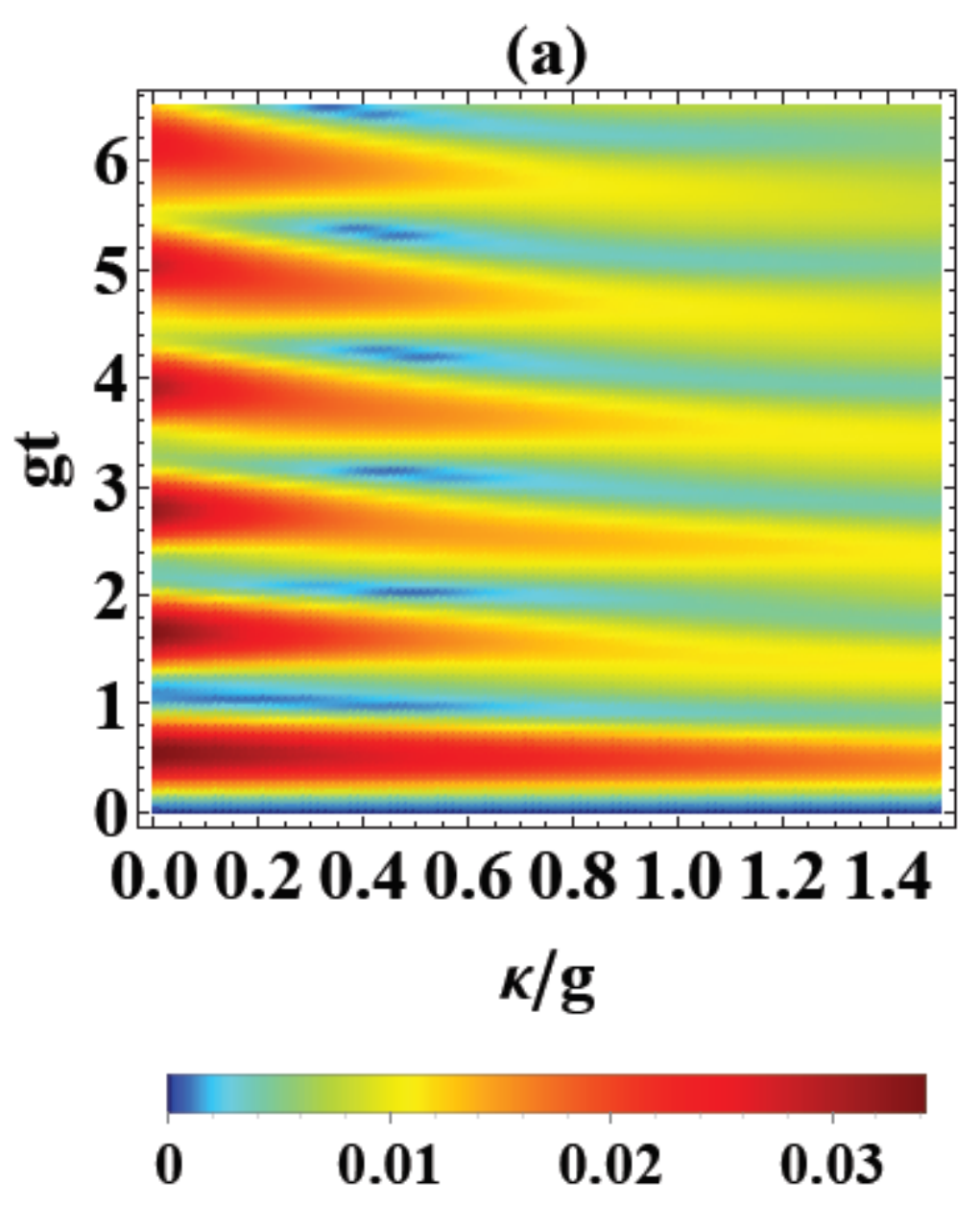}} & 
\subfloat{\includegraphics[width=5.5cm,height=6cm]{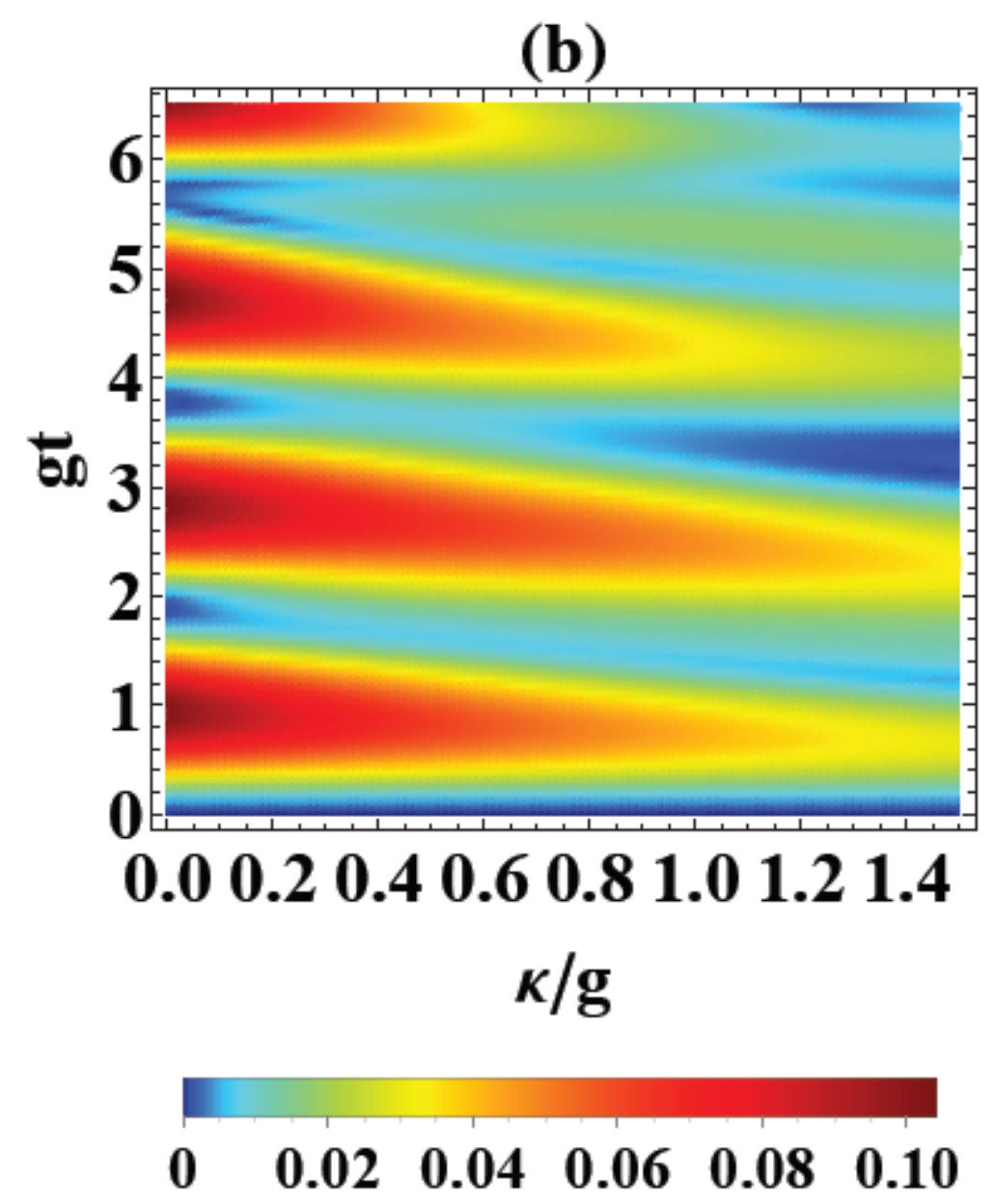}}&
\subfloat{\includegraphics[width=5.5cm,height=6cm]{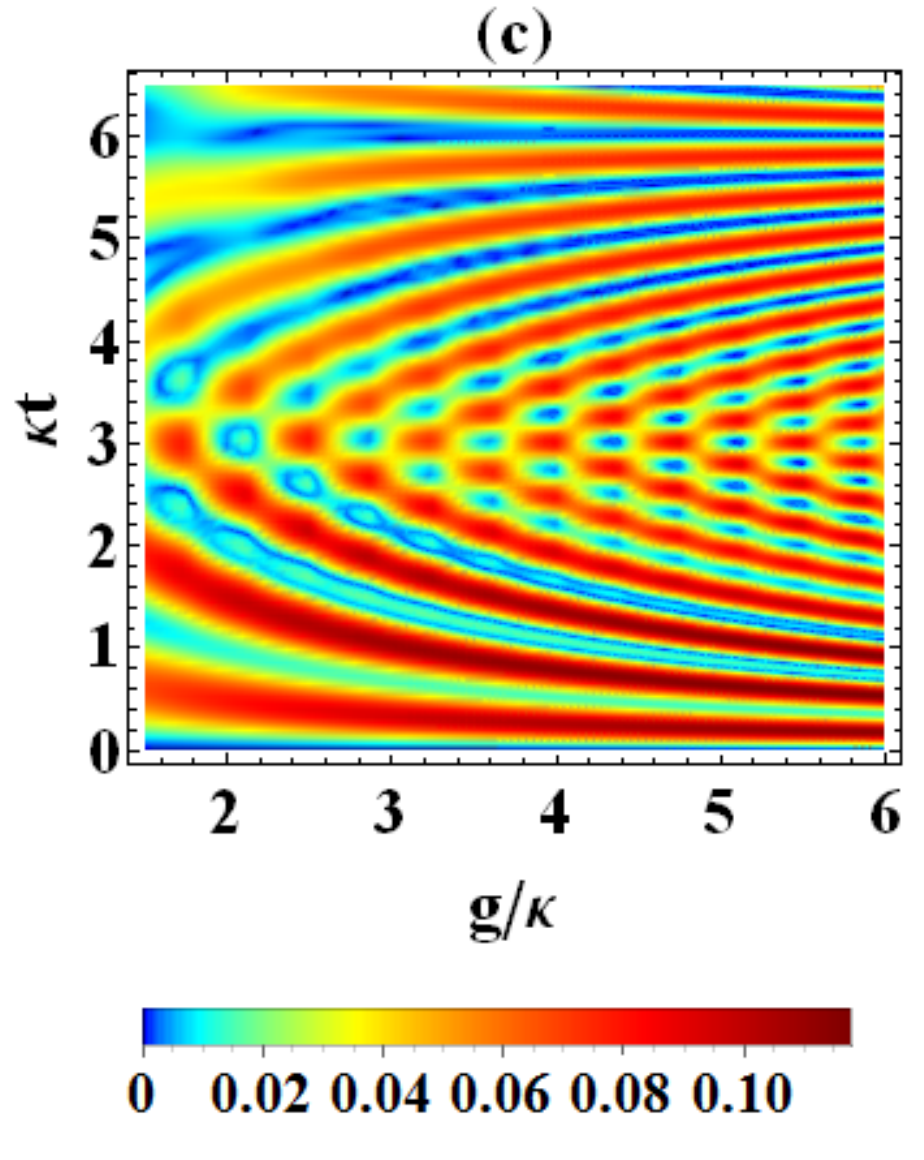}}\\
\end{tabular}
\captionsetup{
  format=plain,
  margin=1em,
  justification=raggedright,
  singlelinecheck=false
}
\caption{Two-photon N00N state generation and evolution in scheme-I. All emitter-cavity system parameters are chosen to be identical for simplicity and particular attention has been paid to (a) large atom-cavity detuning (b) cavity mode backscattering and (c) stronger coupling regime. The rest of the parameters are the same as used in Fig.~\ref{Fig4}.}\label{Fig5}
\end{center}
\end{figure*}
\subsection{Bi-photon NOON state}
After the discussion of single-photon N00N state (Bell states), in this section, we focus on the two-photon N00N states of the form
\begin{align}
\ket{\Psi}=\frac{1}{\sqrt{2}}(\ket{2,0}+\ket{0,2}).    
\end{align}
In literature, this state is also known as the Hong-Ou-Mandel state (after the well-known 1987 Hong-Ou-Mandel experiment \cite{hong1987measurement}) and has its significance in the context of pure quantum interference effects among indistinguishable photons \cite{di2014observation, lewis2014proposal,lopes2015atomic,branczyk2017hong}. At the two-photon (and later for a higher number of photon N00N states) our scheme-I and scheme-II start to differ. Therefore, we'll discuss both schemes now separately.


\subsubsection{Setup for Scheme-I}
In this scheme, we'll consider four emitter-cavity subsystems in the JC array where we divide the system into two groups. The first two emitter-cavity systems form one group and the last two emitter-cavity systems form the other. Initially, we'll start in two excitations N00N state constructed through QEs, while considering either the first group having both QEs excited and other group's QEs in the ground state or vice versa. Again, our key interest is to generate a photonic N00N state in cavities and examine the parameter regimes where both the highest fidelity and longer survival time can be realized. To this end, we rewrite the master equation in a more suggestive form (using the general form of the equation worked out in Eq.~\eqref{mas-(1)}) and identifying the following set of jump/output operators
\begin{subequations}
\begin{eqnarray}
\hat{J}_{e}=\sum^{8}_{j=2,4}\sqrt{\kappa_{j}}\hat{a}_{j},\\
\hat{J}_{o}=\sum^{7}_{j=1,3}\sqrt{\kappa_{j}}\hat{a}_{j}.
\end{eqnarray}
\end{subequations}
The no-jump state ($\ket{\tilde{\psi}}$) has now seventy-four different possibilities of distributing two excitations among twelve slots in the kets (four QEs and eight intra-cavity modes). Still, one can work out the ensemble-averaged density operator which symbolically appears as
\begin{equation}
\hat{\rho}(t)=\ket{\tilde{\psi}}\bra{\tilde{\psi}}+|c_{g}(t)|^{2}\ket{\varnothing}\bra{\varnothing},
\end{equation}
while $\ket{\varnothing}$ is the ground state for the full system (all QEs unexcited and no photons in the cavity modes).

Following the same line of calculations adopted for the case of single-photon N00N state case, we trace out the emitter part from $\hat{\rho}_s(t)$ and obtain $\hat{\rho}_c(t)$. Using $\hat{\rho}_c(t)$, in Fig.~\ref{Fig5} we numerically plot the two-photon N00N state fidelity for the present scheme. We notice an overall marked decrease in the two-photon N00N state fidelity when compared to the single-photon case, but an overall enhanced storage ability (in all three parts of Fig.~\ref{Fig5} the required state is trapped longer than $6$ time units). Fig.~5\ref{Fig5}(a) shows an almost order of magnitude reduction in $\mathcal{F}_{c}$ (compared to Bell-state case) but larger amplitude oscillations in the fidelity when a large emitter-cavity detuning is introduced. This happens since increasing detuning atoms being far from resonance, doesn't emit the photon quickly into the corresponding cavity modes which result in an overall suppression in the cavity mode-based N00N state. 

\begin{figure*}
\begin{center}
\begin{tabular}{cccc}
\subfloat{\includegraphics[width=5.5cm,height=6cm]{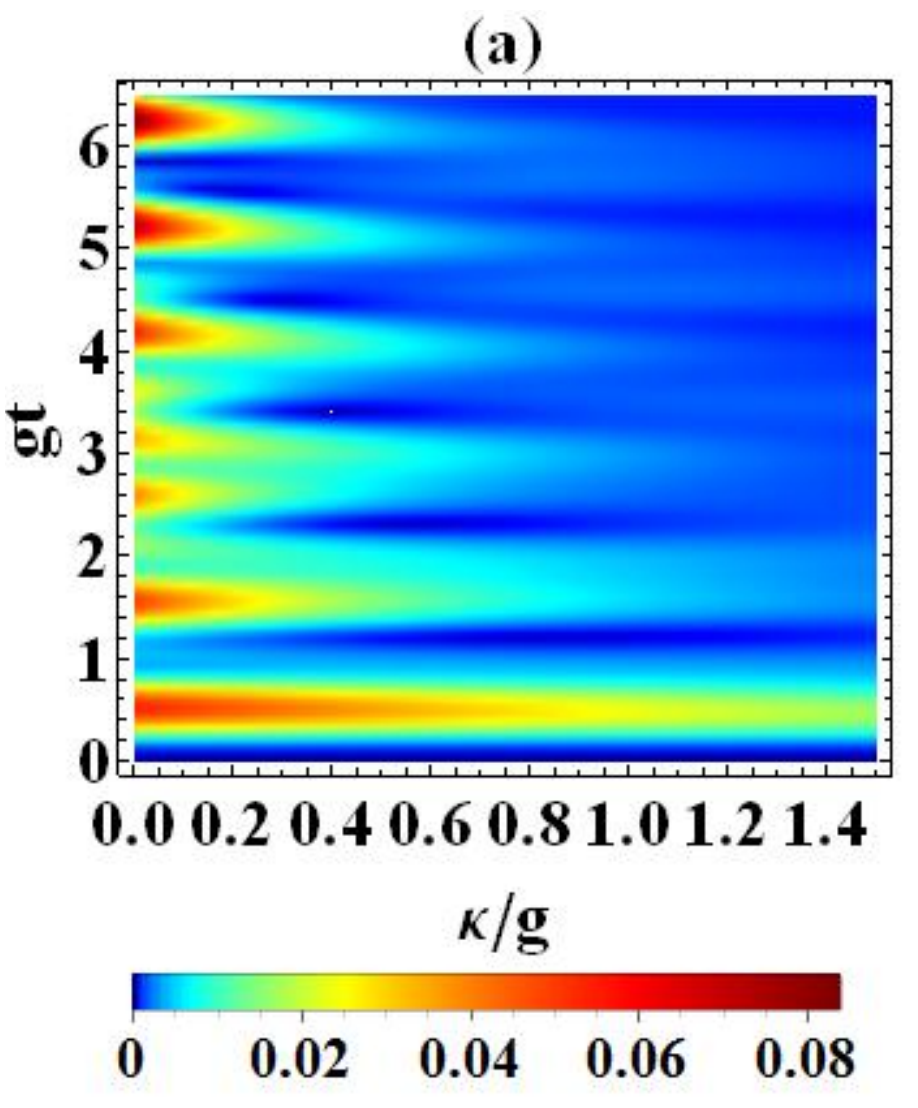}} & 
\subfloat{\includegraphics[width=5.5cm,height=6cm]{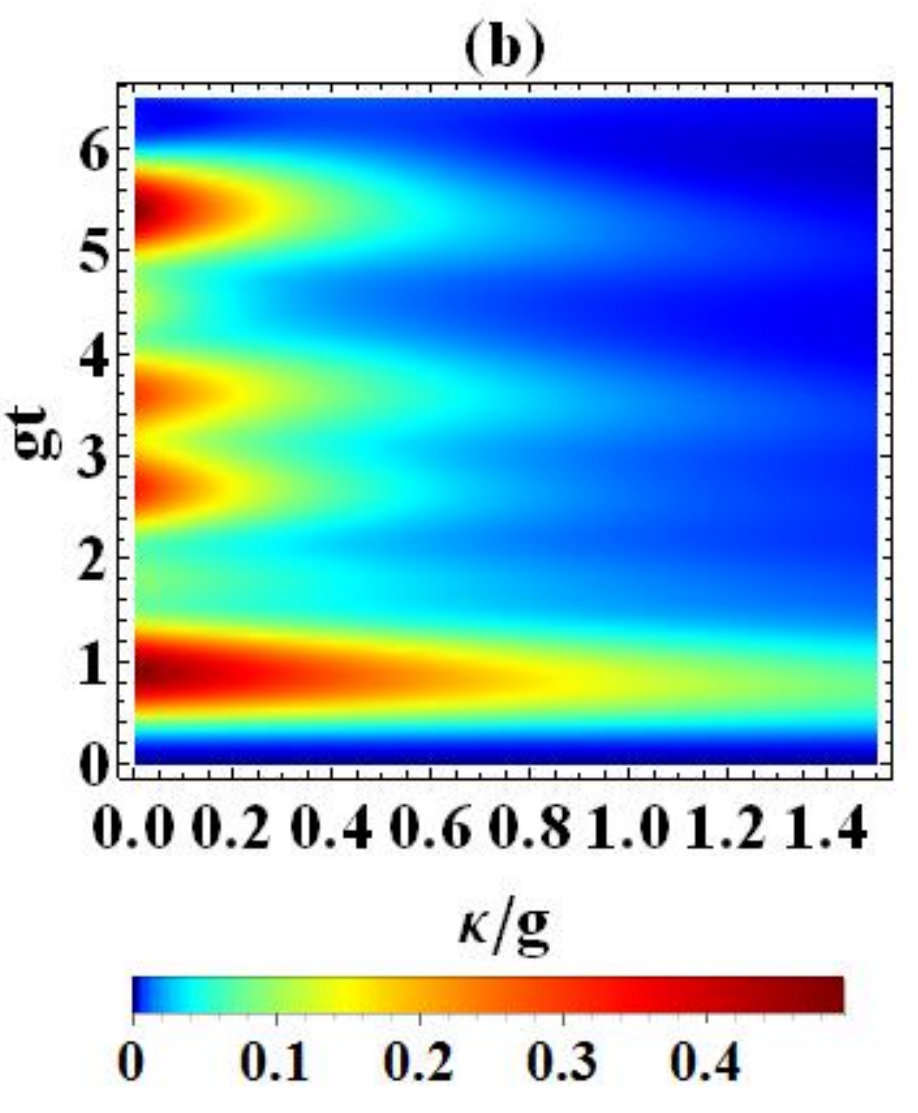}}&
\subfloat{\includegraphics[width=5.5cm,height=6cm]{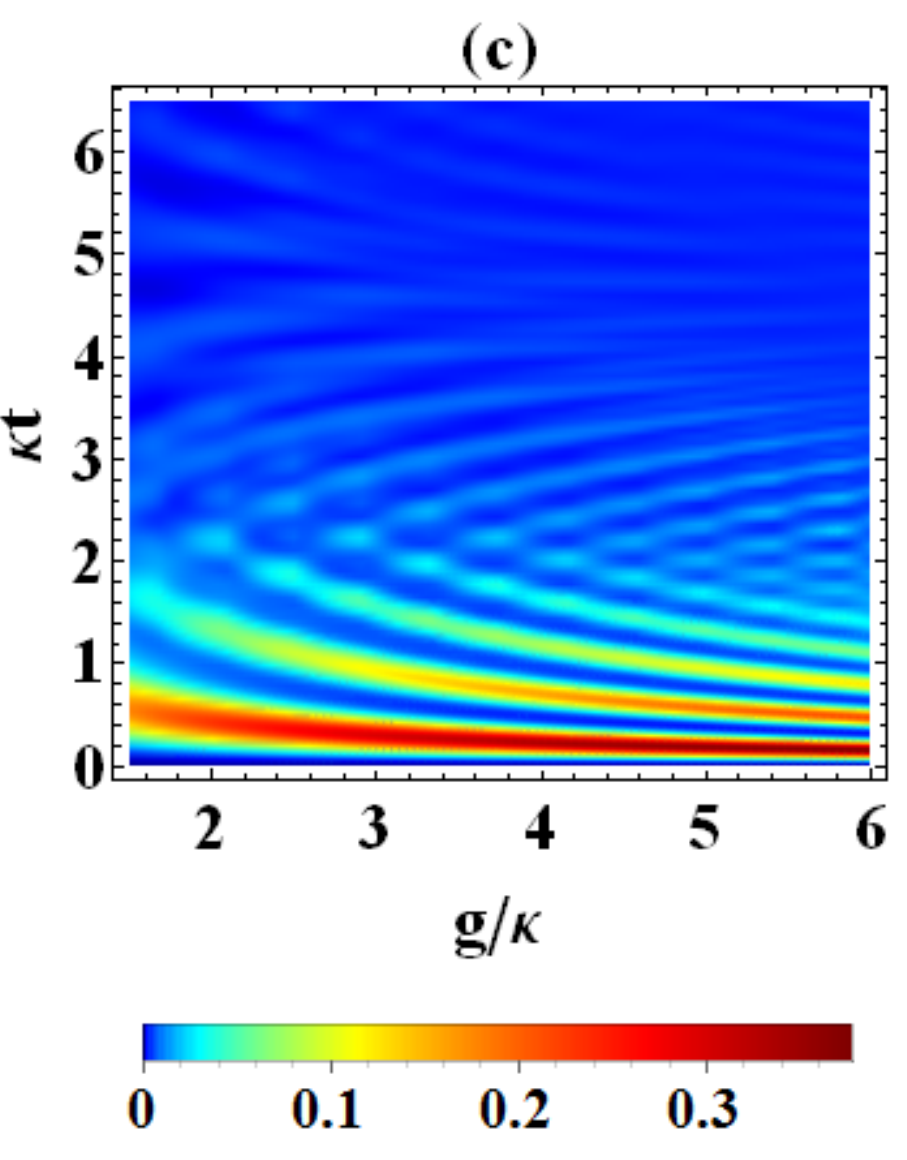}}\\
\end{tabular}
\captionsetup{
  format=plain,
  margin=1em,
  justification=raggedright,
  singlelinecheck=false
}
\caption{Two-photon N00N state decay employing scheme-II. (a) large atom-cavity detuning case, (b) cavity mode backscattering, and (c) Strong coupling regime. In all of the plots, a small DDI interaction $\xi$ has been included between the two QEs. Specifically, $\xi=0.5g$ for (a) and (b) part of the figure and $0.5\kappa$ for the part (c).  The rest of the parameters are the same as used in Fig.~\ref{Fig5}.}\label{Fig6}
\end{center}
\end{figure*}

Next, the introduction of cavity backscattering enhances the fidelity from 3.3\% to 10\% (compare Fig.~5(a) and Fig.~\ref{Fig5}(b)). Again fidelity shows blobs of collapse and revival, but with the passage of time blobs start to separate due to an increase in the destructive interference of many different probability amplitudes pathways in which the two photons can shuttle back and forth in the JC array. Finally, as shown in Fig.~\ref{Fig5}(c), entering a stronger coupling regime causes a slight increase in the maximum fidelity (from 10\% to 13\%) while the two-photon N00N state fidelity does not oscillate in the usual manner rather exhibits a more involved quantum interference pattern.


\subsubsection{Setup for Scheme-II}
In scheme-II we always consider two coupled cavities but for a two-photon state we couple each RR with two dipole-dipole interacting emitters. We assume, initially either both QEs coupled to the left cavity being excited and the emitters coupled to the right cavity are in their ground state or vice versa. The set of jump operators for this case take the form
\begin{subequations}
\begin{eqnarray}
\hat{J}_{o}=\sqrt{\kappa_{1}}\hat{a}_{1}+\sqrt{\kappa_{3}}\hat{a}_{3},\\
\hat{J}_{e}=\sqrt{\kappa_{2}}\hat{a}_{2}+\sqrt{\kappa_{4}}\hat{a}_{4}.
\end{eqnarray}
\end{subequations}
And the no-jump state has thirty-two different ways of finding the system with two excitations before the recording of any photon. Here we explicitly report this state as
\begin{widetext}
\begin{align}
&\ket{\tilde{\psi}}= ~\Big(c_{1}(t)\hat{\sigma}^\dagger_1\hat{\sigma}^\dagger_2+c_{2}(t)\hat{\sigma}^\dagger_1\hat{\sigma}^\dagger_3+c_{3}(t)\hat{\sigma}^\dagger_1\hat{\sigma}^\dagger_4+c_{4}(t)\hat{\sigma}^\dagger_2\hat{\sigma}^\dagger_3+c_{5}(t)\hat{\sigma}^\dagger_2\hat{\sigma}^\dagger_4+c_{6}(t)\hat{\sigma}^\dagger_3\hat{\sigma}^\dagger_4+c_7(t)\hat{\sigma}^\dagger_1\hat{a}^\dagger_1+c_8(t)\hat{\sigma}^\dagger_1\hat{a}^\dagger_2+c_9(t)\hat{\sigma}^\dagger_1\hat{a}^\dagger_3\nonumber\\
&+c_{10}(t)\hat{\sigma}^\dagger_1\hat{a}^\dagger_4+c_{11}(t)\hat{\sigma}^\dagger_2\hat{a}^\dagger_1+c_{12}(t)\hat{\sigma}^\dagger_2\hat{a}^\dagger_2+c_{13}(t)\hat{\sigma}^\dagger_2\hat{a}^\dagger_3+c_{14}(t)\hat{\sigma}^\dagger_2\hat{a}^\dagger_4+c_{15}(t)\hat{\sigma}^\dagger_3\hat{a}^\dagger_1+c_{16}(t)\hat{\sigma}^\dagger_3\hat{a}^\dagger_2+c_{17}(t)\hat{\sigma}^\dagger_3\hat{a}^\dagger_3+c_{18}(t)\hat{\sigma}^\dagger_3\hat{a}^\dagger_4\nonumber\\
&+c_{19}(t)\hat{\sigma}^\dagger_4\hat{a}^\dagger_1+c_{20}(t)\hat{\sigma}^\dagger_4\hat{a}^\dagger_2+c_{21}(t)\hat{\sigma}^\dagger_4\hat{a}^\dagger_3+c_{22}(t)\hat{\sigma}^\dagger_4\hat{a}^\dagger_4+c_{23}(t)\hat{a}^\dagger_1\hat{a}^\dagger_1+c_{24}(t)\hat{a}^\dagger_2\hat{a}^\dagger_2+c_{25}(t)\hat{a}^\dagger_3\hat{a}^\dagger_3+c_{26}(t)\hat{a}^\dagger_4\hat{a}^\dagger_4+c_{27}(t)\hat{a}^\dagger_1\hat{a}^\dagger_2\nonumber\\
&+c_{28}(t)\hat{a}^\dagger_1\hat{a}^\dagger_3+c_{29}(t)\hat{a}^\dagger_1\hat{a}^\dagger_4+c_{30}(t)\hat{a}^\dagger_2\hat{a}^\dagger_3+c_{31}(t)\hat{a}^\dagger_2\hat{a}^\dagger_4+c_{32}(t)\hat{a}^\dagger_3\hat{a}^\dagger_4\Big)\ket{\varnothing}.
\end{align}
\end{widetext}
In the same manner, as before, we numerically calculate and plot the desired fidelity of the two-photon N00N state in Fig.~\ref{Fig6}. When we compare each plot in Fig.~\ref{Fig6} with the corresponding plot in Fig.~\ref{Fig5} we notice an overall trend of increase in the fidelity, but the storage of the required state has been diminished considerably. In Fig.~\ref{Fig6}(a) again large detuning causes survival of two-photon state for a longer time but the maximum fidelity achieved is limited to 8.2\% only which is greater than the corresponding maximum fidelity of $\sim $3\% obtained in the JC array case. Inclusion of the backscattering between cavity modes causes a considerable enhancement in the fidelity (maximum value jumps to $50\%$) but the regular oscillatory pattern (as seen for instance in Fig.~\ref{Fig6}(b)) has been disturbed due to an increase in the number of ways in which the destructive interference among different photonic paths can occur. Finally, in Fig.~\ref{Fig6}(c) we find a wave-like oscillatory profile resembling Fig.~\ref{Fig5}(c) but a decrease in the maximum fidelity (compared to Fig.~\ref{Fig6}(b)) to 37\%. The non-linearities produced by the QEs seem to destroy the two-photon detection at the same detector (Hong-Ou-Mandel interference) in this case, similar to what has been reported in Ref.~\cite{mirza2015nonlinear}.


\section{Summary and Conclusions}
By using the QJA approach combined with the input-output formalism of quantum optics, in this paper we have studied the transfer and decay of multiphoton N00N states in two CQED architectures/schemes. Scheme-I consists of two-way cascaded JC arrays whilst scheme-II makes use of multiple QEs that are DDI and are coupled to two coupled cavities. After describing a general theoretical treatment valid for any number of QEs and RRs, as working examples, we presented and examined the situations of single and two-photon N00N states. For the single-photon N00N state (Bell states), in which case, both schemes are the same, we demonstrated that starting in an emitter Bell state one can transfer this state to cavity modes with the maximum fidelity 96\% and also to the emitter-cavity hybrid system with fidelity 50\%. We then focused on the cavity mode case. There we found that cavity modes backscattering serves as the best option to preserve Bell state with maximum fidelity of $\mathcal{F}_c\sim$ 80\% with up to 6$g^{-1}$ storage time.

For the two-photon N00N state scenario, both setups geometrically become different. We compared the required state fidelities in both schemes and found that while working in the strong coupling regime, the JC array scheme is (overall) better for the longer storage of two-photon state with storage time increased by a factor of $6$ and maximum fidelity achieved up to 12\%. Multiple DDI scheme (scheme-II) is well suited for gaining higher fidelity though (up to 50\% for the backscattering case) but for shorter periods due to unavailability of a chain of JC subsystems.

Finally, we remark that our study has shown that unless clever ways of state manipulation are not utilized this trend of fast decay and lower fidelity values would extend down to N00N states with higher $\mathcal{N}$ values. In this context, there have been proposals reported in the last five years or so (see for instance \cite{su2014fast, xiong2015efficient, qi2020generating}) to achieve higher N00N state fidelities. However, none of those studies have focused on the multi-emitter many RRs cascaded schemes considered in this work. We point out that such many-body schemes have practical relevance to recent advancements in quantum computing (with $50- 100$ qubit number)\cite{preskill2018quantum} and in the development of multiple node quantum communication protocols. We, therefore, leave the task of achieving higher fidelities and retaining N00N states (with $\mathcal{N}>2$) in the presence of spontaneous emission for longer times within the two-way cascaded JC setups considered in this study, as a future possible direction of this work.


\section*{Acknowledgements}
IMM would like to acknowledge the support of the Miami University College of Arts and Science and Physics Department start-up funding. ASC would like to thank National Science Foundation Grant No. 1757575 for financial support through the Miami University Physics Department Research Experience for Undergraduates program.

\bibliographystyle{ieeetr}
\bibliography{paper.bib}
\end{document}